\documentclass[useAMS,usenatbib,referee,11pt]{article}

\usepackage[figuresright]{rotating}
\usepackage{natbib}
\usepackage{amssymb,amsmath}
\usepackage{bm,float}

\usepackage{graphicx}
\usepackage[nolists]{endfloat} 
\usepackage{threeparttable} 

\usepackage{epsfig}
\usepackage{latexsym}
\usepackage{amsmath,amssymb,amsfonts,amsxtra,bm}
\usepackage{graphicx,color,pstcol}
\usepackage{natbib}
\usepackage{euscript,mathrsfs,afterpage}
\usepackage[figuresright]{rotating}
\usepackage{multicol,ragged2e}
\usepackage{setspace}
\usepackage{rotating}

\usepackage[title]{appendix}

\newcommand{\cov}{\mbox{cov}}

\vspace{-1in}
\vspace{-.5 in}

\title{	  \vspace{-.2 in}
Regression calibration to correct correlated errors in outcome and exposure \vspace{-.05 in}}

\author{Pamela A. Shaw$^{*}$ \\
	  Department of Biostatistics and Epidemiology, Perelman School of Medicine, \\
	  University of Pennsylvania, Philadelphia, Pennsylvania, USA \\
	  email: shawp@upenn.edu
	  \and 
	  Jiwei He \\
	  Department of Biostatistics and Epidemiology, Perelman School of Medicine, \\
	  University of Pennsylvania, Philadelphia, Pennsylvania, USA  \\
	  email:  jiweihe@upenn.edu
	   \and
	   Bryan E. Shepherd \\
	    Department of Biostatistics, Vanderbilt University School of Medicine, \\
	    Nashville, Tennessee, USA \\
	    email: bryan.shepherd@vanderbilt.edu
	   }

\begin{document}

\maketitle

\date{}




\label{firstpage}

\thispagestyle{empty}       
\newpage
\thispagestyle{empty}       

\begin{abstract}
Measurement error arises through a variety of mechanisms. A rich literature exists on the bias introduced by covariate measurement error and on methods of analysis to address this bias. By comparison, less attention has been given to errors in outcome assessment and non-classical covariate measurement error. We consider an extension of the regression calibration method to settings with errors in a continuous outcome, where the errors may be correlated with prognostic covariates or with covariate measurement error. This method adjusts for the measurement error in the data and can be applied with either a validation subset, on which the true data are also observed (e.g., a study audit), or a reliability subset, where a second observation of error prone measurements are available. For each case, we provide conditions under which the proposed method is identifiable and leads to unbiased estimates of the regression parameter. When the second measurement on the reliability subset has no error or classical unbiased measurement error, the proposed method is unbiased even when the primary outcome and exposures of interest are subject to both systematic and random error.  We examine the performance of the method with simulations for a variety of measurement error scenarios and sizes of the reliability subset. We illustrate the method's application using data from the Women's Health Initiative Dietary Modification Trial.

\noindent \textbf{Keywords:}
Bias; Linear regression; Measurement error; Nutrition Assessment;  Nutritional epidemiology; Regression calibration.

\end{abstract}

\clearpage
\restylefloat{table}



\addtocounter{page}{-2}

\section{Introduction}

Measurement error arises in many biomedical settings. For example, data can be recorded with error in large clinical databases. Errors are also common in exposures that are primarily measured by self-report, such as dietary intakes. These errors can occur in the outcome or covariates, or both. Furthermore, errors in the outcome may be correlated with important prognostic variables or with errors in covariates. While much attention in the literature has focussed on the setting of classical covariate measurement error, less attention has been paid to errors in the outcome. One reason for this is that for independent, mean zero random error in the outcome, the usual linear regression parameter estimates are unbiased; however, if the error in the outcome is related to prognostic variables or the errors in the covariates, then the estimated regression coefficients will be biased. Some work has been done in the setting of covariate-independent error in the response for the generalized linear model \citep{buona96} and several have considered methods to address misclassification in the response \citep{pepe92,magder97, meier03, kuchenhoff06, edwards13}. Our interest is the linear regression setting where there may be error in both the response and covariates.

One setting where errors are common in both outcome and exposures is large clinical studies, particular those using data primarily collected for non-research purposes (e.g., administrative databases and electronic health records).  Certain types of records (e.g. records from a particular study site) are often more likely to have errors, records with errors often tend to have errors across multiple variables, and the magnitude of these errors may be correlated. Given correlation between the errors in the outcome and exposure covariates, ignored errors in the data could lead to conservative or non-conservative bias in estimates of regression parameters. In some studies, data validation or audits are performed in subsets of records for quality control purposes. Although often discarded, findings from audit data can be used to adjust analyses for errors which remain in the unaudited data. We will consider a measurement error correction method for linear regression that produces unbiased estimates in the presence of correlated errors using audit/validation data. \cite{shepherd11} proposed a moment correction method for linear regression for this setting and applied their method to data from an HIV observational cohort study. \cite{shepherd12} considered similar methods in the context of audited randomized clinical trials. We consider an extension of the popular regression calibration method to this setting. 

We also address a setting not  considered by previous work to address correlated errors in outcomes and covariates, one where no validation subset is available, but where a second, objective measurement, i.e. with unbiased classical measurement error, for the error prone exposures and outcome is available on a subset. In epidemiology, there are many examples of exposures and outcomes that are self-reported, and subject to both random and systematic error, but for which an objective measurement can also be obtained by a more expensive or more invasive procedure. One such setting comes from nutritional epidemiology, where patterns of dietary consumption are of interest; namely how one dietary attribute, such as energy intake, is associated with other dietary intakes, all of which are measured by self-reported questionnaire data. A common instrument to measure dietary intake, the food frequency questionnaire, has been shown to have both systematic and random error \citep{willet}; and errors in outcomes and exposures of interest assessed by the FFQ are likely correlated.  Social pressure to lead a healthy lifestyle could correlate the systematic measurement error across self-reported outcomes and exposures.  For some nutrients, there are objective biomarkers, known as recovery markers, which capture actual intake up to mean zero, random error. Two such examples are the doubly-labeled water recovery marker for total energy consumption \citep{schoeller88} and a 24 hour urinary nitrogen recovery marker for protein intake \citep{Bingham85}. Due to expense and participant burden, a large prospective cohort study generally could only include these recovery markers on a small subset.  We show that our methods can also be applied in this setting, where the primary exposure and outcome measures are subject to errors which are correlated and may consist of both systematic bias and random error and where an objective measure for these variables is available in a sub-cohort. 

In this paper we develop an extension of regression calibration that yields unbiased parameter estimates in linear regression in the presence of systematic and random measurement error in the outcome and covariates, as well as potentially correlated errors between the outcome and covariates. Regression calibration, introduced by \cite{prentice82}, has become a popular method for addressing covariate measurement error, likely because of its easy implementation and its good numerical performance in a broad range of settings \citep{carroll06}. Most regression calibration methods assume classical, unbiased measurement error in the covariate only, and to date these methods have not addressed correlated errors between outcomes and exposures. We consider the case where the correlated errors are mean zero and where, such as in the FFQ example, errors in both outcome and covariates could be biased. We provide assumptions under which repeat measures on a subset are sufficient for the proposed method and that the true outcomes and covariates do not need to ever be observed. We also consider the case where the true data are observed in a subset.

We examine the numerical performance of the proposed method for a variety of measurement error scenarios and compare its performance to the naive solution that ignores the error. We then illustrate the method with a data example using nutritional assessment data from the Women's Health Initiative Dietary Modification Trial \citep{whi}.

\section{Measurement Error Model} \label{errormodel}
Let $Y$ be a continuous outcome, $X$ a $p \times 1$ vector of covariates that may be observed with error, and $Z$ a $q \times 1$ vector of accurately observed covariates. Let $X^\star$ and $Y^\star$ be error prone versions of $X$ and $Y$, respectively. Assume $(Y,X,Z)$ follows the linear model
\begin{equation} \label{model}
Y = \beta_0 + \beta_x' X + \beta_z' Z + \epsilon,
\end{equation} 
where $\epsilon$ is mean zero random error that is independent of all other random variables in equation (\ref{model}). Instead of observing $(X, Z,Y)$, $(X^\star,Z, Y^\star)$  are observed on a cohort of $N$ $iid$ individuals and a random subset of $n$ individuals have been selected to have a second measure of $X$ and $Y$. We consider three cases: 1) where the second observation is a repeat observation of $(X^\star,Y^{\star})$, 2) where the second measure is the true $(X,Y)$, and 3) where the second measure is a different error prone measure of $(X,Y)$, namely $(X^\star_B,Y^\star_B)$, which is an objective biomarker measure whose errors are mean zero and independent of all other variables. Let $V$ be the indicator that an individual is selected to have a second measure of $(X,Y)$. In the measurement error setting, this subset is referred to as a reliability subset when a second error prone measure is observed and a validation subset when the truth observed. We will refer to the group of individuals for which $V=1$ as the reliability subset in case 1;  as the validation or audit subset in case 2, and the biomarker subset in case 3. We next define the notation and measurement error model for each of these cases.

\subsection{Case 1: Reliability subset}
Define
\begin{equation} \label{rely}
\begin{split}
 X_{ij}^\star  & = X_i + T_{ij} \\
 Y_{ij}^\star   &= Y_i + \tilde{T}_{ij}, 
 \end{split}
\end{equation}
\noindent where $i=1,\ldots,N$; $j=1,\ldots,k_i$; and $k_i = 2$ for individuals where $V_i=1$ and 1 otherwise. (This could be easily extended to $k_i>2$ for some individuals.) Suppose $(T_{ij},\tilde{T}_{ij})$ are random error terms independent and identically distributed across individuals and with mean zero. We allow $\cov(T_{ij},\tilde{T}_{ij}) \ne 0$, but we assume that the error terms in repeat observations of $X_{ij}^{\star}$ and $Y_{ij}^{\star}$ are independent, namely $\cov(T_{ij},\tilde{T}_{ij'}) =\cov( \tilde{T}_{ij},\tilde{T}_{ij'}) = \cov(T_{ij},{T}_{ij'}) = 0$ for $j \ne j'$. In this case, where only repeat measures of $(X^\star,Y^\star)$ are available, we must also assume that we have random error for the necessary nuisance parameters in the error model to be identifiable. That is, for case 1 we have the additional assumption that the error terms $T$ and $\tilde{T}$ are independent of $(X,Y,Z)$. For cases 2 and 3, we can relax this assumption.

\subsection{Case 2: Validation subset} 
Here we assume the same additive error model as in case 1, only that $k_i=1$ for all subjects and for the subset of $n$ individuals where $V_i=1$, we assume the true covariate and outcome $(X_i,Y_i)$ are also observed. When discussing this case, we will drop the second subscript $j$ to emphasize there are no repeat observations. Since $X$ and $Y$ are observed in a subset of individuals, we can also allow for a more general error model, including differential error. That is, we can allow $(T_i,\tilde T_i)$ to have nonzero mean and nonzero covariance with $(X_i,Y_i,Z_i)$ without losing identifiability.

A motivating setting for the validation subset case is that of the data audit of clinical studies discussed by \cite{shepherd11}. In this setting, it is assumed that a subset of individuals are selected for a data audit that will ascertain the true outcome and exposure. These authors considered a mixture model for  $(T_i,\tilde T_i)$ with point mass at 0. We consider this model as a special case of equation (\ref{rely}).

\subsection{Case 3: Biomarker subset}
Motivated by our data example, we consider the more general error model that allows the errors $(T_{i},\tilde{T}_{i})$, in addition to being potentially correlated, to have nonzero mean due to potential scale and location bias in $(X_i^\star,Y_i^\star)$. We assume a general measurement error model for the setting of nutritional intake data \citep{prentice13}, which has also been applied to physical activity data \citep{neuhouser13}.

We assume that $k_i=1$ for all subjects and that the error for the primary measures $(X_i^\star,Y_i^\star)$ has systematic bias that is a linear function of $Z_i$. Specifically, for $(X_i^\star,Y_i^\star)$ defined by equation $(\ref{rely})$, let
\begin{equation} \label{ffq}
\begin{split}
 T_i & = \alpha_0 + \alpha_1 Z_i+ U_i \\
  \tilde T_i  & = \gamma_0 + \gamma_1 Z_i + \tilde U_i,  
 \end{split}
\end{equation}
where $(U_i,\tilde U_i)$ are potentially correlated mean zero random error terms that are independent of the other terms in the model. For a random subset of $n$ individuals ($V_i=1$), we assume the objective biomarker measurements $(X^\star_{B},Y^\star_{B})$ are also available, which obey a classical error model.  Namely,
\begin{equation} \label{biom}
\begin{split}
 X_{Bi}^\star  & = X_i + \eta_i \\
 Y_{Bi}^\star   &= Y_i + \nu_i,
 \end{split}
\end{equation}
\noindent where that $\eta_i$ and $\nu_i$ are mean zero random errors that are mutually independent as well as independent of all terms in equation (\ref{model}). Because the errors $(\eta,\nu)$ are independent of $(X,Y$), this case will have the same flexibility as case 2, and we can allow $\cov(X,T)$ and $\cov(Y,{\tilde{T}})$ to be nonzero and still have all parameters identifiable on the biomarker subset. Consequently, our proposed methods will be able to correct for a more general measurement error model for the primary measures $(X^\star,Y^\star)$ observed on the whole cohort even when the true $(X,Y)$ are never observed.

\section{Proposed Method}

Following the general approach of regression calibration, we model the expected value of the unobserved data as a function of the observed data and use these quantities to estimate the regression parameters of interest. Using the additivity of the error, one has
\begin{equation*}
\begin{split}
E(Y^\star| X^{\star},Z) &= E(Y|X^{\star},Z) + E(\tilde{T}|X^{\star},Z) \\
&=  E\{E( Y|X,X^{\star},Z)|X^{\star},Z \}+  E(\tilde{T}|X^{\star},Z) \\
&= E(\beta_0 + \beta_x X + \beta_z Z | X^{\star},Z) + E(\tilde{T}| X^{\star},Z) \\
&= \beta_0 + \beta_x E(X|X^{\star},Z) + \beta_z Z + E(\tilde{T} | X^{\star},Z).
\end{split} 
\end{equation*}

\noindent  The third equality holds because, by assumption, $X^\star$ provides no further information about $Y$ than is contained in $(X,Z)$.  This suggests that one could regress $Y^\star - \hat{E}(\tilde{T}|X^{\star},Z)$ on $\{\hat{E}(X|X^{\star},Z), Z\}$ and get an unbiased estimate for $\mathbf{\beta}=(\beta_0,\beta_x,\beta_z)$, where $\hat{E}(\cdot)$ denotes an estimate of the expectation.   If the error term in $Y^\star$ is independent of that in $X^\star$, or when there is no measurement error in the observed outcome variable, one can perform the regression of $Y^\star$ on $\{\hat{E}(X |X^{\star},Z),Z\}$ instead of $(X,Z)$, which is the usual regression calibration approach \citep{prentice82}, to obtain an unbiased estimator of $\mathbf{\beta}=(\beta_0,\beta_x,\beta_z).$ Estimation of  $\hat{E}(\tilde{T}|X^{\star},Z)$ and $\hat{E}(X |X^{\star},Z)$ is described below for each case.

For an estimator of $E(\tilde{T}|X^{\star},Z)$, one can consider the following first order approximation

\begin{equation} \label{estttilde}
E[ \tilde{T} | X^{\star},Z] \dot{=}  \; \mu_{\tilde{T}} + \left [ \begin{matrix}\Sigma_{ \tilde{T}X^{\star}} \; & \Sigma_{\tilde{T}Z}\end{matrix} \right ]  \left [ \begin{matrix} \Sigma_{X^\star} \; & \Sigma_{X^{\star}Z} \;  \\ \Sigma_{ZX^{\star}}\; & \Sigma_{Z} \end{matrix} \right ] ^{-1}  \left [ \begin{matrix} X^\star- \mu_{X^{\star}} \\ Z- \mu_Z  \end{matrix} \right],
\end{equation}
\noindent where $\Sigma_{ab} \equiv \mbox{Cov}(a,b)$ and $\Sigma_{a} \equiv \mbox{Var}(a)$.  Similarly, we approximate $E[ X | X^{\star},Z]$ as 


\begin{equation}\label{estX}
E[ X  | X^{\star},Z]  \dot{=}  \mu_X +  \left [ \begin{matrix} \Sigma_{XX^{\star}} \; & \Sigma_{XZ}\end{matrix} \right ]  \left [ \begin{matrix} \Sigma_{X^\star} \; & \Sigma_{X^{\star}Z} \;  \\ \Sigma_{ZX^{\star}}\; & \Sigma_{Z} \end{matrix} \right ] ^{-1} \left [ \begin{matrix} X^\star- \mu_{X^{\star}} \\ Z- \mu_Z  \end{matrix}\right].
\end{equation}



We first consider case 1, the reliability subset model shown in  equation (\ref{rely}).  In this case, the errors $(T,\tilde{T})$ are independent of $(X,Y,Z)$ and centered at 0.   One can use the following equalities to obtain the moments involving random variables not directly observed: $\mu_{\tilde{T}} = 0$, $\mu_{X} = \mu_{X^\star}$, $\Sigma_{ \tilde{T}X^{\star}} =  \Sigma_{\tilde{T}T} =\cov(Y_1^\star - Y_2^\star,X^\star_{1} - X^\star_2)/2$, $\Sigma_{XX^{\star}} = \Sigma_X= \Sigma_{X^{\star}} - \Sigma_T, \Sigma_T =\mbox{var}(\tilde{X}_{1} - \tilde{X}_{2})/2$, $\Sigma_{\tilde{T}Z} = 0$, and $\Sigma_{XZ}= \Sigma_{X^\star Z}$.  Estimates of ($\Sigma_{\tilde{T}X^{\star}}, \Sigma_T,\Sigma_{XX^{\star}})$ can be estimated using the reliability subset, and estimates of $(\mu_{X^\star},\Sigma_{X^{\star}},\mu_Z,\Sigma_Z,\Sigma_{ZX^{\star}})$ can be obtained from the entire study cohort. 

For case 1, we have assumed subjects in the reliability subset have 2 measures of $(X^\star,Y^\star)$, and those not in the reliability subset have only one measure. In the above formulas one can think of $X^\star$ as a vector of 2 observations or a single observation as appropriate. The Appendix provides an explicit estimating equation for each of the nuisance parameters, which can also be used to provide a sandwich estimator of the variance for the regression parameters of interest using the stacked estimating approach outlined by \cite{stefanski02}. 

Case 2 also uses equations (\ref{estttilde}) and (\ref{estX}) for estimation of $E[ X  | X^{\star},Z] $ and $E[ \tilde{T} | X^{\star},Z]$. For case 2, all necessary moments in equations (\ref{estttilde}) and (\ref{estX}) can be directly estimated, since both $(X,Y)$ and $(X^\star,Y^\star)$ are observed from the validation subset. This allows for a more general error model to be identifiable with the data. For example, $T$ and $\tilde{T}$ could be allowed to be correlated with $Z$. In this case, we need to estimate $\Sigma_{\tilde{T}Z}$ and $\Sigma_{XZ}$ from the validation subset. The stacked estimation equations used for the M-estimator and the sandwich estimator of the variance for this case are also provided in the Appendix. 

For case 3, we consider an error model that allows for systematic and correlated error, but for which the true $(X,Y)$ can never be observed. From equation (\ref{ffq}), one has
\begin{eqnarray*}
E(Y^\star| X^\star,Z) &=& \beta_0 + \beta_x E(X|X^\star,Z) + \beta_z  Z + c^\star, 
\end{eqnarray*}

\noindent where $c^\star =E(\gamma_0 + \gamma_1Z + \tilde U | X^\star,Z)$, which has a different functional form than in cases 1 and 2. One can once again estimate $\beta$ by regressing $Y^\star - \hat c^\star$ on $\{\hat E(X|X^{\star},Z),Z\}$. In this case, the parameters necessary to calculate $\hat c^\star$ are estimated by regressing $Y^\star - Y_B$ on $(Z,X^\star)$ in the biomarker subset. One could also choose to estimate $E(X|X^{\star},Z)$ with a similar regression approach. Because the biomarker $X_B$ only involves classical measurement error, $E(X_B|X^{\star},Z)=E(X|X^{\star},Z)$ and the nuisance parameters necessary for $\hat E(X|X^{\star},Z)$ can be estimated by the regression \[ E[X_B |X^\star, Z] = \alpha_0 + \alpha_x X^\star + \alpha_z Z. \]

The stacked estimation equations for the sandwich estimator of the variance for case 3 are provided in the Appendix. The standard errors for the proposed method can also be calculated using the bootstrap for each of the cases studied. 

 
\section{Simulation Study} \label{simulations}

We examine the finite sample performance of the proposed regression calibration method using numerical simulation. In this section we focus on case 1, assuming a reliability subset is available. We examine a limited set of simulations for case 2, given the similarity with case 1. Numerical simulations for case 3 are presented in the next section as part of the data example.  For the different parameter scenarios for the error and outcome regression models, we compare the numerical performance of the proposed method with that of standard linear regression using the true $(X,Y)$ (\textit{True} method) and the error prone observed data $(X^\star,Y^\star)$ without correction (\textit{Naive} method).    For all scenarios, we summarize results across 1000 Monte Carlo simulations. 

We assume the linear regression model $Y = \beta_0 + \beta_x X + \beta_z Z + \epsilon$, where $\mu_X = 0$, $\sigma_X=1$, and $\sigma_\epsilon = 5 $, where $\sigma_X$ is the standard deviation for $X$ and similarly for $\epsilon$. For Scenario 1, we simulate the simple linear regression model $(\beta_z=0)$, with $\beta_0=2$ and let $\beta_x \in \{1,5\}$. For Scenario 2, we consider the ANCOVA model and additionally study the effect of measurement error on a precisely observed covariate $Z$, setting $(\beta_0,\beta_x,\beta_z) = (2,1,-1)$, $\sigma_z=1$, and $\rho_{xz} = \mbox{cor(X,Z)} \in \{0,0.5\}$. For the measurement error parameters, we consider $\sigma_T \in \{0.5,1\}$, assume $\sigma_{\tilde T}= \sigma_{T}$ and allow $\rho_{T \Tilde{T}}\equiv\mbox{cor}(T,\tilde T)$  to vary, with $\rho_{T {\tilde T}} \in \{-0.5,-0.25,0,0.25, 0.5\}$.  Note the size of the error was chosen to represent moderate error, where the variance of the error is 25\% the variance of the true exposure $X$, and large error, where the error variance is equal to the variance of $X$.   For Scenarios 1 and 2 we consider normally distributed data, with a total cohort size of N=400 and we assume we have a reliability subset of size $n$, for which a second measure of $(X^\star,Y^\star)$ is available. We vary the size of this reliability subset, with $n \in \{25,50,100,200,400\}$. We then consider simulations with $N=1000$, as well as simulations with non-Gaussian error distributions.

Table 1 presents the results for Scenario 1, with $\beta_x=1$. For all the parameter configurations in this table, there is appreciable bias in the \textit{Naive} estimates, \textit{i.e.} in the estimate based on error prone $(X^\star,Y^\star)$ with no adjustment for the measurement error. Coverage is generally poor for the \textit{Naive} estimator, with worse performance for the larger measurement error and the low to negative correlations between $\tilde T$ and $T$ ($\rho_{T{\tilde T}}$).   The relative performance of the proposed method depended on the size of the reliability subset. Some small sample bias was present for $n=25$, but for $n=50$ and larger there was a notable improvement in all scenarios, with bias generally between 1-5\% and diminishing  to less than 1\% for larger $n$. The mean squared error (MSE) for the proposed method was generally competitive or smaller than that for the \textit{Naive} estimator for $n=50$ and a marked improvement for larger $n$, with one exception being that for the smaller $\sigma_T=\sigma_{\tilde T}=0.5$ and high positive correlation the \textit{Naive} estimator maintained the smallest MSE for all $n$. The average standard error estimated from the M-estimation (ASE) compared well to the empirical standard error (SE) and provided good nominal coverage. The coverage probability (CP) for the proposed method was generally in the range of 93-95\% and comparable to the CP of 93.8\% for the \textit{True} estimator from regression using $(X,Y)$.

\begin{table}[h!b!p!]
\scriptsize
\caption[width=.2 in]{For 1000 simulated data sets  of size N=1000, the mean percent (\%) bias, empirical standard
error (SE), average estimated standard error (ASE), mean squared error
(MSE) and 95\% coverage probability (CP) are given for $\beta_x$. Results are provided for linear regression using the true data $(X,Y)$ (TRUE), the error prone data $(X^\star,Y^\star)$ (NAIVE), and the proposed $(\hat{X},\hat{Y})$ estimated using a reliability subset (denoted by size of reliably subset $n=25,50,200,400$). Data are generated according to Scenario 1 in Section \ref{simulations}, with $\beta_x=1$.}
\begin{center}
\begin{tabular}{lccccccccccc}
\hline
\hline
 & & \multicolumn{5}{c}{$\sigma_T=\sigma_{\tilde T}=1$} & \multicolumn{5}{c}{$\sigma_T=\sigma_{\tilde T}=0.5$}  \\
\cline{3-7}
\cline{8-12}
   $\rho_{T \tilde T}$ & Method & \% Bias & SE & ASE & MSE & CP & \% Bias & SE & ASE & MSE & CP \\
  \hline
  
    0.5 & TRUE & -1.10 & 0.26 & 0.25 & 0.07 & 0.938 & -1.10 & 0.26 & 0.25 & 0.07 & 0.938 \\ 
   & NAIVE & -25.58 & 0.19 & 0.18 & 0.10 & 0.699 & -10.93 & 0.23 & 0.23 & 0.07 & 0.909 \\ 
   & 25 & 2.56 & 0.79 & 0.98 & 0.62 & 0.958 & -1.03 & 0.31 & 0.29 & 0.09 & 0.934 \\ 
   & 50 & 0.79 & 0.44 & 0.42 & 0.20 & 0.949 & -1.11 & 0.31 & 0.29 & 0.10 & 0.928 \\ 
   & 100 & -0.85 & 0.41 & 0.38 & 0.17 & 0.939 & -1.40 & 0.31 & 0.29 & 0.10 & 0.928 \\ 
   & 200 & -0.52 & 0.37 & 0.35 & 0.13 & 0.945 & -1.04 & 0.30 & 0.28 & 0.09 & 0.934 \\ 
   & 400 & -0.61 & 0.32 & 0.31 & 0.11 & 0.939 & -0.93 & 0.28 & 0.26 & 0.08 & 0.937 \\ 
   \\
  0.25 & TRUE & -1.10 & 0.26 & 0.25 & 0.07 & 0.938 & -1.10 & 0.26 & 0.25 & 0.07 & 0.938 \\ 
   & NAIVE & -38.08 & 0.19 & 0.18 & 0.18 & 0.455 & -15.92 & 0.24 & 0.23 & 0.08 & 0.879 \\ 
   & 25 & 4.05 & 1.14 & 1.43 & 1.30 & 0.951 & -0.89 & 0.31 & 0.29 & 0.10 & 0.930 \\ 
   & 50 & 1.95 & 0.47 & 0.45 & 0.22 & 0.946 & -1.01 & 0.31 & 0.29 & 0.10 & 0.929 \\ 
   & 100 & -0.27 & 0.42 & 0.39 & 0.17 & 0.943 & -1.32 & 0.31 & 0.29 & 0.10 & 0.934 \\ 
   & 200 & -0.27 & 0.37 & 0.36 & 0.14 & 0.950 & -1.00 & 0.30 & 0.29 & 0.09 & 0.934 \\ 
   & 400 & -0.49 & 0.33 & 0.31 & 0.11 & 0.933 & -0.90 & 0.28 & 0.26 & 0.08 & 0.938 \\ 
   \\
  0 & TRUE & -1.10 & 0.26 & 0.25 & 0.07 & 0.938 & -1.10 & 0.26 & 0.25 & 0.07 & 0.938 \\ 
   & NAIVE & -50.58 & 0.19 & 0.18 & 0.29 & 0.217 & -20.92 & 0.24 & 0.23 & 0.10 & 0.833 \\ 
   & 25 & 8.99 & 1.07 & 0.97 & 1.14 & 0.945 & -0.74 & 0.32 & 0.30 & 0.10 & 0.929 \\ 
   & 50 & 3.20 & 0.50 & 0.48 & 0.25 & 0.943 & -0.91 & 0.31 & 0.29 & 0.10 & 0.931 \\ 
   & 100 & 0.35 & 0.43 & 0.40 & 0.18 & 0.946 & -1.23 & 0.31 & 0.29 & 0.10 & 0.933 \\ 
   & 200 & -0.01 & 0.38 & 0.36 & 0.14 & 0.940 & -0.95 & 0.30 & 0.29 & 0.09 & 0.933 \\ 
   & 400 & -0.37 & 0.33 & 0.32 & 0.11 & 0.934 & -0.88 & 0.28 & 0.26 & 0.08 & 0.940 \\ 
   \\
  -0.25 & TRUE & -1.10 & 0.26 & 0.25 & 0.07 & 0.938 & -1.10 & 0.26 & 0.25 & 0.07 & 0.938 \\ 
   & NAIVE & -63.09 & 0.19 & 0.18 & 0.43 & 0.073 & -25.92 & 0.24 & 0.23 & 0.12 & 0.775 \\ 
   & 25 & 18.31 & 2.47 & 2.38 & 6.12 & 0.939 & -0.58 & 0.32 & 0.30 & 0.10 & 0.931 \\ 
   & 50 & 4.58 & 0.54 & 0.51 & 0.29 & 0.940 & -0.80 & 0.32 & 0.30 & 0.10 & 0.932 \\ 
   & 100 & 1.02 & 0.44 & 0.42 & 0.19 & 0.950 & -1.14 & 0.32 & 0.29 & 0.10 & 0.933 \\ 
   & 200 & 0.26 & 0.38 & 0.37 & 0.15 & 0.944 & -0.91 & 0.30 & 0.29 & 0.09 & 0.936 \\ 
   & 400 & -0.25 & 0.33 & 0.32 & 0.11 & 0.934 & -0.86 & 0.28 & 0.27 & 0.08 & 0.942 \\ 
   \\
  -0.5 & TRUE & -1.10 & 0.26 & 0.25 & 0.07 & 0.938 & -1.10 & 0.26 & 0.25 & 0.07 & 0.938 \\ 
   & NAIVE & -75.6 & 0.19 & 0.18 & 0.61 & 0.020 & -30.92 & 0.24 & 0.23 & 0.15 & 0.714 \\ 
   & 25 & 9.03 & 2.92 & 3.10 & 8.53 & 0.928 & -0.43 & 0.33 & 0.31 & 0.11 & 0.934 \\ 
   & 50 & 6.12 & 0.59 & 0.55 & 0.35 & 0.940 & -0.69 & 0.32 & 0.30 & 0.10 & 0.939 \\ 
   & 100 & 1.72 & 0.45 & 0.43 & 0.21 & 0.947 & -1.04 & 0.32 & 0.30 & 0.10 & 0.935 \\ 
   & 200 & 0.53 & 0.39 & 0.38 & 0.15 & 0.947 & -0.87 & 0.30 & 0.29 & 0.09 & 0.938 \\ 
   & 400 & -0.14 & 0.33 & 0.32 & 0.11 & 0.936 & -0.84 & 0.28 & 0.27 & 0.08 & 0.942 \\ 
  \hline
    \end{tabular}
  \end{center}
\end{table}
\normalsize

Table 2 presents results for the same set of parameters, except now letting $\beta_x=5$. The performance of the regression calibration estimator had little small sample bias and was comparatively unaffected by the size of $\beta$. The coverage probability was much poorer for the \textit{Naive} estimator, and the proposed method maintained good coverage and the smallest MSE for all scenarios with $n>50$ and for the smaller error variance also had the smallest MSE for $n=25$.

  \begin{table}[h!b!p!]
\scriptsize
\caption[width=.2 in]{For 1000 simulated data sets  of size N=1000, the mean percent (\%) bias, empirical standard
error (SE), average estimated standard error (ASE), mean squared error
(MSE) and 95\% coverage probability (CP) are given for $\beta_x$. Results are provided for linear regression using the true data $(X,Y)$ (TRUE), the error prone data $(X^\star,Y^\star)$ (NAIVE), and the proposed $(\hat{X},\hat{Y})$ estimated using a reliability subset (denoted by size of reliably subset $n=25,50,200,400$). Data are generated according to Scenario 1 in Section \ref{simulations}, with $\beta_x=5$.}
\begin{center}
\begin{tabular}{lccccccccccc}
\hline
\hline
 & & \multicolumn{5}{c}{$\sigma_T=\sigma_{\tilde T}=1$} & \multicolumn{5}{c}{$\sigma_T=\sigma_{\tilde T}=0.5$}  \\
\cline{3-7}
\cline{8-12}
   $\rho_{T{\tilde T}}$ & Method & \% Bias & SE & ASE & MSE & CP & \% Bias & SE & ASE & MSE & CP \\
  \hline
 0.5 & TRUE & -0.22 & 0.26 & 0.25 & 0.07 & 0.938 & -0.22 & 0.26 & 0.25 & 0.07 & 0.938 \\ 
   & NAIVE & -45.06 & 0.21 & 0.21 & 5.12 & 0.000 & -18.19 & 0.25 & 0.24 & 0.89 & 0.034 \\ 
   & 25 & 7.58 & 7.25 & 7.82 & 52.64 & 0.869 & -0.16 & 0.44 & 0.43 & 0.20 & 0.939 \\ 
   & 50 & 2.73 & 1.14 & 1.08 & 1.33 & 0.901 & -0.22 & 0.39 & 0.37 & 0.15 & 0.939 \\ 
   & 100 & 0.70 & 0.75 & 0.72 & 0.56 & 0.939 & -0.34 & 0.36 & 0.34 & 0.13 & 0.933 \\ 
   & 200 & 0.59 & 0.55 & 0.54 & 0.31 & 0.950 & -0.15 & 0.33 & 0.31 & 0.11 & 0.937 \\ 
   & 400 & 0.23 & 0.44 & 0.42 & 0.19 & 0.932 & -0.15 & 0.30 & 0.28 & 0.09 & 0.937 \\  \\
   
  0.25 & TRUE & -0.22 & 0.26 & 0.25 & 0.07 & 0.938 & -0.22 & 0.26 & 0.25 & 0.07 & 0.938 \\ 
   & NAIVE & -47.55 & 0.22 & 0.22 & 5.70 & 0.000 & -19.19 & 0.25 & 0.24 & 0.98 & 0.026 \\ 
   & 25 & 3.21 & 9.36 & 10.12 & 87.57 & 0.865 & -0.09 & 0.46 & 0.45 & 0.21 & 0.937 \\ 
   & 50 & 3.12 & 1.20 & 1.14 & 1.47 & 0.906 & -0.16 & 0.40 & 0.38 & 0.16 & 0.936 \\ 
   & 100 & 0.86 & 0.78 & 0.75 & 0.61 & 0.938 & -0.3 & 0.36 & 0.34 & 0.13 & 0.937 \\ 
   & 200 & 0.63 & 0.57 & 0.56 & 0.33 & 0.948 & -0.14 & 0.33 & 0.32 & 0.11 & 0.939 \\ 
   & 400 & 0.23 & 0.45 & 0.43 & 0.20 & 0.934 & -0.15 & 0.30 & 0.28 & 0.09 & 0.933 \\ \\
   
  0 & TRUE & -0.22 & 0.26 & 0.25 & 0.07 & 0.938 & -0.22 & 0.26 & 0.25 & 0.07 & 0.938 \\ 
   & NAIVE & -50.06 & 0.22 & 0.22 & 6.31 & 0.000 & -20.19 & 0.25 & 0.25 & 1.08 & 0.021 \\ 
   & 25 & 7.38 & 4.99 & 3.90 & 25.08 & 0.870 & -0.01 & 0.48 & 0.46 & 0.23 & 0.934 \\ 
   & 50 & 3.56 & 1.28 & 1.21 & 1.66 & 0.909 & -0.10 & 0.41 & 0.39 & 0.17 & 0.927 \\ 
   & 100 & 1.07 & 0.81 & 0.78 & 0.66 & 0.935 & -0.26 & 0.37 & 0.35 & 0.13 & 0.937 \\ 
   & 200 & 0.68 & 0.59 & 0.58 & 0.35 & 0.949 & -0.13 & 0.33 & 0.32 & 0.11 & 0.941 \\ 
   & 400 & 0.23 & 0.46 & 0.44 & 0.21 & 0.933 & -0.15 & 0.30 & 0.28 & 0.09 & 0.936 \\ \\
   
  -0.25 & TRUE & -0.22 & 0.26 & 0.25 & 0.07 & 0.938 & -0.22 & 0.26 & 0.25 & 0.07 & 0.938 \\ 
   & NAIVE & -52.57 & 0.23 & 0.22 & 6.96 & 0.000 & -21.19 & 0.26 & 0.25 & 1.19 & 0.014 \\ 
   & 25 & 12.37 & 8.12 & 7.76 & 66.31 & 0.869 & 0.08 & 0.50 & 0.48 & 0.25 & 0.931 \\ 
   & 50 & 4.09 & 1.37 & 1.28 & 1.93 & 0.910 & -0.04 & 0.42 & 0.40 & 0.18 & 0.929 \\ 
   & 100 & 1.31 & 0.85 & 0.82 & 0.72 & 0.940 & -0.21 & 0.37 & 0.35 & 0.14 & 0.933 \\ 
   & 200 & 0.74 & 0.61 & 0.59 & 0.37 & 0.946 & -0.12 & 0.34 & 0.32 & 0.11 & 0.942 \\ 
   & 400 & 0.24 & 0.47 & 0.45 & 0.22 & 0.937 & -0.15 & 0.30 & 0.29 & 0.09 & 0.940 \\ \\
   
  -0.5 & TRUE & -0.22 & 0.26 & 0.25 & 0.07 & 0.938 & -0.22 & 0.26 & 0.25 & 0.07 & 0.938 \\ 
   & NAIVE & -55.09 & 0.23 & 0.23 & 7.64 & 0.000 & -22.2 & 0.26 & 0.25 & 1.30 & 0.010 \\ 
   & 25 & 5.36 & 10.10 & 10.34 & 102.1 & 0.861 & 0.17 & 0.52 & 0.49 & 0.27 & 0.931 \\ 
   & 50 & 4.71 & 1.49 & 1.36 & 2.29 & 0.907 & 0.03 & 0.43 & 0.41 & 0.19 & 0.928 \\ 
   & 100 & 1.60 & 0.88 & 0.85 & 0.79 & 0.941 & -0.15 & 0.38 & 0.36 & 0.14 & 0.932 \\ 
   & 200 & 0.81 & 0.63 & 0.61 & 0.39 & 0.939 & -0.10 & 0.34 & 0.32 & 0.12 & 0.941 \\ 
   & 400 & 0.26 & 0.47 & 0.46 & 0.22 & 0.940 & -0.14 & 0.30 & 0.29 & 0.09 & 0.938 \\
  \hline
    \end{tabular}
  \end{center}
\end{table}
\normalsize

In Table 3, we show the results for the ANCOVA model for a similar set of measurement error parameters as in Table 1 and 2, fixing $\beta_x=1$ and focusing on positive correlation for $\rho_{T{\tilde T}}$. As expected, the \textit{Naive} estimator for $\beta_z$ is unaffected by the measurement error in $X^\star$ for the scenarios where $\rho_{xz}=0$ \citep{shepherd12}. For the other scenarios  there is bias in the \textit{Naive} estimator and lower than the nominal 95\% coverage for both $\beta_z$ and $\beta_x$. The regression calibration estimates perform well across the different parameter choices for the measurement error model, with good coverage and the small sample bias diminishing with increasing size of the reliability subset. It is notable that in these simulations, when $X$ is correlated with $Z$ ($\rho_{xz}=0.5$) and $\rho_{T{\tilde T}} \ne 0$,  the MSE for the regression calibration estimator is larger than that for the \textit{Naive} estimator except with the larger reliability subsets. This is due to the increased uncertainty in the regression calibration estimates for $\beta$.  Supplementary Table S1 in the Web Appendix examines scenarios similar to Table 3, only with smaller measurement error variance parameters $\sigma_T= \sigma_{\tilde T}=0.5$.  We see that for the \textit{Naive} estimator, there is less bias and smaller MSE for both $\beta_x$ and $\beta_z$ with the smaller measurement error variance; however the \textit{Naive} estimator has low coverage due to its bias for both $\beta_x$ and $\beta_z$, whereas the proposed method again maintains close to 95\% coverage in all scenarios.  Supplementary Table S2 presents results for cohort size N=1000 and reliability subset sizes $n=50, 100, 200, 500, 1000$.  The performance of the RC estimator improves with the larger reliability sizes, as expected. The coverage of the \textit{Naive} estimator gets worse, as this estimator becomes more certain about the wrong thing. Supplementary Table S3 considers the effects of negative correlation for the error terms $T$ in $X^\star$ and $\tilde T$ in $Y^\star$, allowing $\rho_{T{\tilde T}} = -0.25, -0.5$.  Patterns in this case are similar to Table 3. The \textit{Naive} estimator had poor coverage; whereas the proposed method maintained good coverage and generally a similar or better MSE for reliability subsets of size 50 or larger, with percent bias diminishing $n$ increased. Supplementary Table S4 considered the relative performance for larger values of $\beta$. Similar to the results in Table 2, the Naive estimator had larger bias and lower coverage; whereas the the proposed method maintained good coverage and for scenarios with $n>25$ generally had competitive or better MSE. The proposed method had larger MSE and an inflated estimate of SE for the small reliability subset size $n=25$.


\begin{table}
\scriptsize
  \renewcommand{\thetable}{\arabic{table}}
\caption{For 1000 simulated data sets of size N=1000, the mean percent (\%) bias, empirical standard
error (SE), average estimated standard error (ASE), mean squared error
(MSE) and 95\% coverage probability (CP) are given for $(\beta_x,\beta_z)$. Results are provided for linear regression using the true data $(X,Z,Y)$ (TRUE),  $(X^\star,Z,Y^\star)$ (NAIVE), and the proposed $(\hat{X},Z,\hat{Y})$ estimated with a reliability subset (Method denoted by size of reliably subset $n=25,50,200,400$). Data are generated according to Scenario 2 in Section \ref{simulations}, with $\beta_x=1$, $\beta_z=-1$ and $\sigma_T=\sigma_{\tilde T}=1$.}
\begin{tabular}{lcccccccccccccccccccc}
\hline
\hline
&  & & \multicolumn{5}{c}{$\beta_x$} & \multicolumn{5}{c}{$\beta_z$}  \\
\cline{4-8}
\cline{9-13}
  $\rho_{xz}$ & $\rho_{T{\tilde T}}$ & Method & \% Bias & SE & ASE & MSE & CP & \% Bias & SE & ASE & MSE & CP \\ 
  \hline
 0.5 & 0.5 & TRUE & -1.05 & 0.29 & 0.29 & 0.08 & 0.955 & 0.39 & 0.29 & 0.29 & 0.08 & 0.949 \\ 
   &  & NAIVE & -29.39 & 0.19 & 0.19 & 0.12 & 0.686 & 14.61 & 0.27 & 0.27 & 0.09 & 0.913 \\ 
   &  & 25 & 1.81 & 3.04 & 6.51 & 9.25 & 0.970 & -1.19 & 1.36 & 3.00 & 1.85 & 0.967 \\ 
   &  & 50 & 2.90 & 0.63 & 0.63 & 0.40 & 0.967 & -1.59 & 0.41 & 0.42 & 0.17 & 0.960 \\ 
   &  & 100 & 1.11 & 0.50 & 0.49 & 0.25 & 0.959 & -0.84 & 0.37 & 0.37 & 0.14 & 0.945 \\ 
   &  & 200 & -0.01 & 0.44 & 0.44 & 0.20 & 0.945 & 0.02 & 0.35 & 0.35 & 0.12 & 0.945 \\ 
   &  & 400 & -0.41 & 0.38 & 0.38 & 0.15 & 0.952 & 0.01 & 0.32 & 0.32 & 0.10 & 0.955 \\ \\
   
   & 0.25 & TRUE & -1.05 & 0.29 & 0.29 & 0.08 & 0.955 & 0.39 & 0.29 & 0.29 & 0.08 & 0.949 \\ 
   &  & NAIVE & -43.59 & 0.20 & 0.19 & 0.23 & 0.393 & 21.73 & 0.27 & 0.27 & 0.12 & 0.886 \\ 
   &  & 25 & 5.93 & 1.56 & 2.94 & 2.43 & 0.959 & -3.35 & 0.79 & 1.45 & 0.63 & 0.962 \\ 
   &  & 50 & 5.02 & 0.64 & 0.64 & 0.41 & 0.962 & -2.64 & 0.41 & 0.43 & 0.17 & 0.960 \\ 
   &  & 100 & 2.33 & 0.53 & 0.51 & 0.28 & 0.957 & -1.43 & 0.38 & 0.38 & 0.15 & 0.945 \\ 
   &  & 200 & 0.47 & 0.45 & 0.45 & 0.20 & 0.947 & -0.21 & 0.35 & 0.35 & 0.12 & 0.942 \\ 
   &  & 400 & -0.22 & 0.38 & 0.39 & 0.15 & 0.948 & -0.08 & 0.32 & 0.32 & 0.10 & 0.955 \\ \\
   
   & 0 & TRUE & -1.05 & 0.29 & 0.29 & 0.08 & 0.955 & 0.39 & 0.29 & 0.29 & 0.08 & 0.949 \\ 
   &  & NAIVE & -57.78 & 0.20 & 0.19 & 0.37 & 0.139 & 28.85 & 0.27 & 0.28 & 0.16 & 0.817 \\ 
   &  & 25 & 25.41 & 3.61 & 6.84 & 13.07 & 0.946 & -13.09 & 1.92 & 3.66 & 3.70 & 0.961 \\ 
   &  & 50 & 7.48 & 0.68 & 0.68 & 0.47 & 0.955 & -3.87 & 0.43 & 0.45 & 0.19 & 0.961 \\ 
   &  & 100 & 3.47 & 0.55 & 0.53 & 0.30 & 0.961 & -1.99 & 0.39 & 0.38 & 0.15 & 0.944 \\ 
   &  & 200 & 0.93 & 0.46 & 0.46 & 0.21 & 0.946 & -0.42 & 0.36 & 0.35 & 0.13 & 0.942 \\ 
   &  & 400 & -0.04 & 0.39 & 0.39 & 0.15 & 0.947 & -0.17 & 0.32 & 0.32 & 0.10 & 0.954 \\ \\
   
  0 & 0.5 & TRUE & -0.91 & 0.24 & 0.25 & 0.06 & 0.955 & 0.10 & 0.25 & 0.25 & 0.06 & 0.947 \\ 
   &  & NAIVE & -25.76 & 0.18 & 0.18 & 0.10 & 0.728 & 0.14 & 0.25 & 0.26 & 0.06 & 0.950 \\ 
   &  & 25 & 3.35 & 0.59 & 0.59 & 0.35 & 0.964 & 0.18 & 0.26 & 0.27 & 0.07 & 0.946 \\ 
   &  & 50 & 3.00& 0.83 & 0.81 & 0.69 & 0.960 & 0.01 & 0.26 & 0.28 & 0.07 & 0.943 \\ 
   &  & 100 & 0.34 & 0.39 & 0.38 & 0.15 & 0.952 & -0.08 & 0.27 & 0.27 & 0.07 & 0.946 \\ 
   &  & 200 & -0.11 & 0.36 & 0.35 & 0.13 & 0.941 & 0.17 & 0.27 & 0.27 & 0.07 & 0.943 \\ 
   &  & 400 & -0.47 & 0.31 & 0.31 & 0.09 & 0.949 & 0.03 & 0.25 & 0.25 & 0.06 & 0.952 \\ \\
   
   & 0.25 & TRUE & -0.91 & 0.24 & 0.25 & 0.06 & 0.955 & 0.10 & 0.25 & 0.25 & 0.06 & 0.947 \\ 
   &  & NAIVE & -38.18 & 0.18 & 0.18 & 0.18 & 0.450 & 0.14 & 0.25 & 0.26 & 0.06 & 0.950 \\ 
   &  & 25 & 10.71 & 1.53 & 1.86 & 2.34 & 0.956 & 0.42 & 0.27 & 0.31 & 0.07 & 0.948 \\ 
   &  & 50 & -2.17 & 1.42 & 1.50 & 2.01 & 0.956 & 0.22 & 0.27 & 0.30 & 0.07 & 0.946 \\ 
   &  & 100 & 1.01 & 0.40 & 0.39 & 0.16 & 0.953 & -0.07 & 0.27 & 0.27 & 0.07 & 0.944 \\ 
   &  & 200 & 0.19 & 0.36 & 0.36 & 0.13 & 0.942 & 0.16 & 0.27 & 0.27 & 0.07 & 0.943 \\ 
   &  & 400 & -0.35 & 0.31 & 0.31 & 0.10 & 0.950 & 0.02 & 0.25 & 0.25 & 0.06 & 0.952 \\ \\
   
   & 0 & TRUE & -0.91 & 0.24 & 0.25 & 0.06 & 0.955 & 0.10 & 0.25 & 0.25 & 0.06 & 0.947 \\ 
   &  & NAIVE & -50.59 & 0.18 & 0.18 & 0.29 & 0.201 & 0.14 & 0.26 & 0.26 & 0.07 & 0.950 \\ 
   &  & 25 & 4.43 & 2.82 & 4.47 & 7.93 & 0.944 & 0.24 & 0.29 & 0.42 & 0.09 & 0.947 \\ 
   &  & 50 & 0.78 & 0.97 & 0.81 & 0.95 & 0.953 & 0.19 & 0.27 & 0.28 & 0.07 & 0.950 \\ 
   &  & 100 & 1.66 & 0.41 & 0.41 & 0.17 & 0.948 & -0.06 & 0.27 & 0.27 & 0.07 & 0.946 \\ 
   &  & 200 & 0.48 & 0.37 & 0.37 & 0.13 & 0.946 & 0.16 & 0.27 & 0.27 & 0.07 & 0.945 \\ 
   &  & 400 & -0.24 & 0.31 & 0.32 & 0.10 & 0.953 & 0.01 & 0.25 & 0.25 & 0.06 & 0.953 \\ 
    \end{tabular}
\end{table}
\normalsize

\clearpage

For Scenario 3, we simulated non-normal distributions for the error and covariates.  We consider a similar simulation to Scenario 1, letting $\sigma_T=\sigma_{\tilde T}=1$, but consider both a mixture of two normals and a log-normal distribution. Supplementary Figure S1 shows these distributions. Simulation results are presented in Supplementary Table S5. The proposed method still provides estimates with diminishing small sample bias for both distributions as $n$ increased; but for the log-normal errors, performance is noticeably better for the larger reliability subsets. For the highly skewed log-normal distribution, $\hat{\beta}$ estimates for the scenarios with reliability subsets smaller than 200 have some small-sample bias. These were generally caused by a few extreme values across the simulations, and the median estimates were much closer to being unbiased (data not shown). 

Simulations for case 2 are provided in Supplementary Tables S6a and S6b. In addition to the \textit{True} and \textit{Naive} methods, the \textit{Proposed} results are compared to those from the related moment-based correction method by Shepherd and Yu (2011). Parameter values for the simulations for this scenario are chosen from those explored by Shepherd and Yu for direct comparison. The proposed regression calibration and the method of \cite{shepherd11} are both a type of moment correction estimate and are asymptotically equivalent for the linear error models we consider in cases 1 and 2. For the case of simple univariate regression they can be made equivalent by choosing the same estimators of the necessary moments (Table S6a). For multivariate regression, the two methods differ on finite samples but had comparable performance (Table S6b). 

\cite{shepherd11} only considered the setting where there was a validation subset available; however, we extend their method to the setting of case 1, where the true values are never observed on anyone, by using the moment estimators provided in the Appendix for the necessary nuisance parameters. Simulation results comparing regression calibration with the moment correction estimates for case 1 are shown in Supplemental Table S7. We see that the two methods again generally perform well for $n>50$ and provide similar estimates, with low small sample bias and very comparable MSE across the scenarios. There was some instability for $n=25$; there appears to be a slight advantage of smaller MSE for the proposed method over the moment correction method for this small reliability size.

In the next section, we examine the performance of regression calibration for case 3, where the measurements $X^\star$ and $Y^\star$ have systematic bias as well as correlated errors, after considering a motivating data example.

\section{Nutritional Epidemiology Data Example} \label{whi}

Tools for dietary assessment of usual intake rely primarily on self-reported data from instruments like the food frequency questionnaire (FFQ). The FFQ consists of a list of foods and a frequency of response for how often each food was consumed over a specified period, such as the last 3 months, and is translated into specific intakes using a nutrient database.   The FFQ has been shown for many nutrients of interest, such as energy and protein, to contain systematic error associated with subject characteristics (body mass index (BMI), gender, age, etc.), as well as within person variability. Despite these known measurement error problems, the FFQ is the most common diet instrument in large cohort studies \citep{willet} because of its low cost. Measurement error for different nutrient intakes assessed with the same FFQ are likely to be correlated.  We consider this problem examining the association between protein density and total caloric intake. 
We consider data from the usual diet (control) arm of the Women's Health Initiative (WHI) Dietary Modification (DM) Trial, which measured dietary intake on 29,294 women using an FFQ.  Because the baseline FFQ was used to
determine eligibility in the DM trial by requiring a minimum of 32\% estimated calories from
fat, the baseline for this analysis was taken as one year after enrollment, at which time
another FFQ was obtained. The DM trial also included a Nutritional Biomarker Substudy (NBS, N=544). The NBS collected self-reported intake along with several objective biomarkers on randomly selected weight-stable women, including doubly-labeled water and urinary nitrogen, with repeat measures on a subset (n=110).  These biomarkers are considered unbiased for short-term usual intake of energy and protein, respectively.  Results of the NBS study are reported by \cite{neuhouser08}. They found BMI to be a
strong determinant of subject-specific bias, with underreporting of energy and protein intake increasing with increasing BMI.


We consider the regression of the log-transformed energy $(Y)$ on log-transformed protein density $(X)$ and BMI ($Z$). Here we assume we have the measurement error structure of case 3 in Section 2.3, where the main outcome and exposure of interest have both systematic bias and potentially correlated measurement error, along with an objective marker with independent classical measurement error for both these variables on a subset. We compare the regression coefficients for log(protein density) and BMI, $\beta_x$ and $\beta_z$ respectively, for regressions using the naive approach (based on error-prone self-reported FFQ data only) and the proposed regression calibration method. From the naive analysis using data from the entire DM cohort, $(\hat{\beta}_x,\hat{\beta}_z) =(-0.14,0.004) $ $(p<0.001,p<0.001)$.  Applying the proposed method to calibrate the FFQ data on the large cohort, we found $\hat{\beta} = (\hat{\beta}_x,\hat{\beta}_z) =(-0.31,0.012) $ $(p=0.43,p<0.001)$.  The coefficient for protein ($X$) appears de-attenuated but is no longer significant due to a large increase in the standard errors. The coefficient of BMI is also larger compared to the analysis based on self-report, suggesting BMI-related bias in self-reported FFQ data was biasing both coefficients of the regression.

Since the biomarkers for protein and energy obey the classical measurement error model, we could obtain an alternate consistent estimator of $\hat{\beta}$ using a complete case analysis on the NBS subset with covariate-only regression calibration to adjust for the error in the protein biomarker and no necessary adjustments for the energy biomarker as the outcome.   A technique from the survey literature, called raking or survey calibration, can be used to combine the estimators \citep{lumley11,deville93}. This could have two potential benefits. First, because it is combining two estimators there is the potential for efficiency gains. Second, in this real data example, where the exact nature of the systematic error of the FFQ is unknown, applying raking of the proposed method using a consistent estimator on the biomarker subset will remove any bias in the proposed method that may have been caused by an inadequate error model. The \textsf{survey} package in R was used to implement the raking procedure, i.e. combine this classical RC estimator with the proposed \citep{R,lumleyBook}. For ease of computation, standard errors were estimated using the bootstrap method.  Using regression calibration on the biomarker subset one has  $(\hat{\beta}_x,\hat{\beta}_z) =(-0.34,0.009) $ $(p=0.07,p<0.001)$. The raking estimator $(\hat{\beta}_x,\hat{\beta}_z) =(-0.35,0.009) $ $(p=0.07,p<0.001)$ was successful in increasing the precision; however, despite the large sample size for our auxiliary estimating equation involving the FFQ, the results were only a modest decrease in the confidence intervals, suggesting that the error-prone FFQ provided only a small amount of information regarding the association between the two dietary intakes.

To better understand the performance of the proposed method under the measurement error structure seen in the WHI example, we built a simulation model on the WHI data. We generated 1000 simulated WHI DM cohorts ($N=29,000$), that each included a biomarker subset ($n=540$). The Supplementary Materials Web Appendix provides further details of the parameters used for this simulation and how they were identified from the WHI data.  In short, for the main cohort and biomarker subset, a regression model for energy as a function of protein density and BMI, as well as measurement error parameters for the observed self-reported FFQ and biomarker versions of protein density and energy, were fit using the observed data. Correlation between the errors of the self-reported protein density and energy are identifiable using the biomarker data. A linear model for the dependence of the error on BMI was assumed and fit to the data. The resulting de-attenuated strength of association between the true $X$ and $Y$ was a similar magnitude for this simulation and the WHI data. The simulation also reproduced the large attenuation seen in the BMI coefficient in the naive analysis of the self-reported data.

Table 4 compares the estimates of $\beta$ for the regression of log energy (Y) on log protein density (X) and BMI (Z) for the regressions based on: the true $(X,Y)$, the error-prone self-reported estimates $(X^{\star},Y^{\star})$, and the proposed regression calibration method incorporating the biomarker observations $(X_B,Y_B)$. We also consider the results using just the regression of $Y_B$ on $(X_B,Z)$ on the biomarker subset, ignoring and accommodating for classical measurement error in $X_B$.  In this case, since the error in both biomarker observations for $Y$ and $X$ follows the classical measurement error model, the error in $Y$ can be ignored and the usual calibration of only $X$ provides an unbiased estimate of $\beta$ in the biomarker subset. The error in the biomarkers for these nutrients are generally considered smaller than that for self-report, so this analysis addresses the question of whether, despite the smaller sample, this analysis may perform better in terms of MSE relative to the proposed and naive analysis. Standard errors for the regression calibration methods were obtained via the bootstrap with 500 bootstrap samples.

The simulation shows the proposed method was close to unbiased, with a mean bias of -1.3\% for $\hat \beta_x$ and $0.30\%$ for $\hat \beta_z$, but it also had large MSE, particularly for $\beta_x$ due to the amount of uncertainty around the slope parameters. The bootstrapped SE were in good agreement with the empirical standard errors. The relative performance of the methods for the simulated analyses in Table 4, in terms of the relative size of the standard errors for the $\beta$ estimates across the methods, is similar to that seen in the data analysis.  Overall, these simulations show that for the more general measurement error model discussed in case 3, the proposed method had good coverage and produced estimates with small bias. They also provide additional evidence that the magnitudes of the error seen in the WHI nutrient data could have led to confidence intervals that included zero despite a true underlying relationship between the nutrients of interest. The naive analysis (ignoring measurement error) had large bias in the regression coefficient for the precisely observed BMI and overall poor coverage for both slope parameters. The analysis using regression calibration for covariate measurement only in the biomarker subset had  smaller mean squared error than the proposed method calibrating both exposure and outcome implemented without raking.  

 In this example, due to the ease of implementation we used bootstrapped standard errors. We examined the relative performance of our method when using the bootstrap versus the sandwich estimate of the variance more generally. We considered the same scenario as shown in the upper left quadrant of Table 3, where samples sizes were smaller and where we might expect more differences between performance of the two standard error estimation approaches. Results are shown in Supplementary Table S8. 
Generally the bootstrap performed well and matched the empirical SE for audit subset sizes of 200 or larger. For smaller audit subsample sizes, particularly $n=25$ or $50$, the bootstrap subset would occasionally produce a spuriously large estimate, leading to some instability. We note for the data example simulations shown in Table 4, the bootstrap performed well compared to the empirical SE which had large sample sizes for both the cohort and subsample. The sandwich estimate across all scenarios studied.

\begin{table}
\scriptsize
 \renewcommand{\thetable}{\arabic{table}}
\caption{For 1000 simulated data sets,  the mean percent (\%) bias, average estimated standard error (ASE), empirical standard
error (SE), mean squared error
(MSE), and coverage probability for the 95\% confidence intervals (CP) are given for $(\beta_x,\beta_z)$, the linear regression coefficients for protein density (log-scale) and BMI using the unadjusted self-reported data on the large cohort (Naive),  the proposed regression calibration method (Proposed), the unadjusted biomarker values on subset (BM Naive), and a complete case regression calibration method that adjusts for the classical measurement error in the biomarker for log-protein density on the biomarker subset (BM RC). The \textit{True} method are the results based on the $(X,Y,Z)$, i.e. (log-energy, log-protein, BMI), measured without error. }
\begin{center}
\begin{tabular}{lcccccccccccccccccccc}
\hline
\hline
&  \multicolumn{5}{c}{$\beta_x$} & \quad &\multicolumn{5}{c}{$\beta_z$}  \\
\cline{2-6}
\cline{8-12}
 Method & \%Bias & SE & ASE & 100*MSE & CP &  & \%Bias & SE & ASE & 100*MSE & CP \\
 \hline
True &  0.0410  & 0.0062 & 0.0064 & 0.0039 & 0.957 &  &-0.0106 & 0.0002 & 0.0002 & 0.0000 & 0.951  \\
Naive      &  16.2380 & 0.0099 & 0.0099 & 0.1069 & 0.114 &   &-80.1324 & 0.0004 & 0.0004 & 0.0110 & 0.000 \\
Proposed   & -1.2704  & 0.0853 & 0.0806 & 0.7278 & 0.934 &   &0.2963 & 0.0025 & 0.0025 & 0.0006 & 0.944 \\
BM Naive  &  -45.8489  & 0.0363 & 0.0360 & 0.9069 & 0.326 & 	& -0.2303 & 0.0018 & 0.0018 & 0.0003 & 0.959 \\
BM RC  & 3.0050  & 0.0749 & 0.0860 & 0.5642 & 0.936 &  & -0.3457 & 0.0019 & 0.0019 & 0.0003 & 0.945\\
\hline
    \end{tabular}
  \end{center}
\end{table}
\normalsize

\section{Discussion}

In this manuscript we have described a regression calibration approach to account for correlated measurement error in both the outcome and the exposure.  Our approach is flexible, simple to implement and can be applied using reliability or validation samples.  Regression calibration methods are popular in practice, and this extension allows their implementation to an important problem.  Although methods for accounting for measurement error have been widely studied, there is surprisingly little on correlated measurement error on the outcome and exposure.  Correlated measurement error is not uncommon; we have seen it from audits of observational data and in nutritional epidemiology settings.

As with all measurement error methods, there are variance/bias trade-offs with using our new method.  With small validation subsets, the reduction in bias may not outweigh the increased variance in estimates, and in some scenarios, naive estimates may have lower mean-squared error than corrected estimates even with large sample sizes.  In applied settings, raking can be used as a method that could potentially improve efficiency by combining the proposed estimator with that of a consistent estimator on the validation/biomarker subset. These facts may motivate multi-stage sampling: the first sample to be large enough to assess the extent of the measurement error problem and the second to be large enough to obtain sufficiently precise corrected estimates.  The size of the second stage of sampling should be based on the first.  Raking was seen in our data analysis to improve precision. More extensive study of these types of sampling procedures is warranted.

For settings where a validation subset is available, a very general error model can be assumed for the data since complete data on the validation subset can be used to estimate the nuisance parameters; though here we must also assume covariates that determine the biased components of the error are observed and modeled correctly. For settings where error-free covariates cannot be observed the methods can still be applied so long as a second error prone measurement whose errors are independent of the first, is available. It is interesting to note that the errors in observed values $(X^{\star},Y^{\star})$ need not be unbiased (mean zero) so long as a second measurement with independent and unbiased errors is available on at least a subset.  When only repeat measures of the same error-prone instrument is observed, the nuisance parameters necessary for the proposed method are only identifiable if the errors are non-differential. 

Development of these types of methods for other types of outcomes is an important direction for future research. More work is also needed on how best to develop an efficient multi-stage design, as efficiency will be key for these methods to be applied to practically sized validation/reliability subsets without undoing the gains in bias-correction by increasing the variability in estimates.

\section*{Acknowledgements}
The authors would like to thank the investigators of the Women's Health Initiative for the use of their data. The WHI program is funded by the National Heart, Lung, and Blood Institute, National Institutes of Health, U.S. Department of Health and Human Services through contracts HHSN268201100046C, HHSN268201100001C, HHSN268201100002C, \\ HHSN268201100003C, HHSN268201100004C, and 
HHSN271201100004C. The work of Dr. Shaw and Dr. Shepherd was supported in part by NIH grant R01-AI131771 and PCORI grant R-1609-36207.  Dr. Shepherd was also supported by NIH Grants R01AI093234 and U01AI069923. 

\section*{Supplementary Materials}
The Web Appendix referenced in Sections~\ref{simulations} and~\ref{whi} will be made available online at the Journal's website as Supplementary Materials for this article.

\bibliography{AuditRC}
\bibliographystyle{chicago}
\clearpage

\begin{appendices}
\section{Standard Error Estimation} \label{varest}
Standard errors of $\hat{\beta}$ can be computed by applying the M-estimation technique and obtaining the sandwich variance estimate (Stefanski and Boos, 2002). A vector $\psi(\theta)$ of stacked estimating equations are formed for the parameter vector $\theta$, which includes both the parameters of interest $(\beta)$ and the nuisance parameters from the measurement error model. The estimates $\hat{\theta}$ can be obtained by solving the equations $\sum_{i=1}^N \psi_i(\theta)=0$ and a sandwich estimator for the variance of $\hat{\theta}$ is estimated as $V(\hat{\theta}) = A(\hat{\theta})^{-1}B(\hat{\theta}) A(\hat{\theta})^{-T}/N, \mbox{ where}$
\begin{equation*}
A(\hat{\theta}) = -\frac{1}{N} \sum_{i=1}^N \frac{\partial \psi_i(\theta)}{\partial \theta}\,\,\,\,\,\,\text{and}\,\,\,\,\,\,B(\hat{\theta}) = \frac{1}{N} \sum_{i=1}^N \psi_i(\theta)\psi_i(\theta)^T.
\end{equation*}
\noindent This technique incorporates the extra variance in $\hat{\beta}$ due to the uncertainty in the nuisance parameters. The first derivative of $\psi_i(\theta)$ can be computed either directly with the assistance of software (e.g. MATLAB) or by numerical differentiation. The variance of $\hat{\beta}$ corresponds to corresponding submatrix of the $V(\hat{\theta})$ matrix. 

In this section we provide the estimating equation $\psi(\theta)$ for each of the cases considered in Section 3. In the expressions below, the half-vectorization operator \textit{vech(A)}  is used to represent a symmetric matrix $A$ as a vector by stacking the columns of the lower triangular portion of the matrix one below the other. In this manner, we create a vector of unique elements for the nuisance parameters in the model that are variance/covariance matrices. 

For \textbf{case 1}, the reliability subset, one has the parameter vector $\theta = (\beta_0, \beta_x, \beta_z, \mu_{X^\star}, \Sigma_{X^*}, \allowbreak \mu_Z, \allowbreak \Sigma_Z , \Sigma_{X^\star Z}, \Sigma_T, \Sigma_{T\, \tilde{T}})$. From these parameters one can also derive other parameters specified in the calibration equations; namely $\mu_X=\mu_{X^\star}$, $\Sigma_{ZX} = \Sigma_{ZX^\star }$, $\Sigma_{X^\star X} = \Sigma_{X^\star} - \Sigma_T$, and $\Sigma_{X^\star \tilde T} = \Sigma_{T \tilde T}$. We define

\begin{equation*}
  \psi_i(\theta)=\begin{cases}
    (\hat{Y}_{i} - \beta_0 - \beta_x \hat{X}_{i} - \beta_z Z_i)  (1 - V_i) + (\hat{Y}_{i1}^{R} + \hat{Y}_{i2}^{R} - 2\beta_0 - 2\beta_x \hat{X}_{i}^{R} - 2\beta_z Z_i)  V_i\\
    (\hat{Y}_{i} - \beta_0 - \beta_x \hat{X}_{i} - \beta_z Z_i)\hat{X}_{i}  (1 - V_i) + (\hat{Y}_{i1}^{R} + \hat{Y}_{i2}^{R} - 2\beta_0 - 2\beta_x \hat{X}_{i}^{R} - 2\beta_z Z_i)\hat{X}_{i}^{R}  V_i\\
    (\hat{Y}_{i} - \beta_0 - \beta_x \hat{X}_{i} - \beta_z Z_i)Z_i  (1 - V_i) + (\hat{Y}_{i1}^{R} + \hat{Y}_{i2}^{R} - 2\beta_0 - 2\beta_x \hat{X}_{i}^{R} - 2\beta_z Z_i)Z_i  V_i\\
(X_{i1}^* + X_{i2}^*  V_i)/(1+V_i) - \mu_{X^\star}\\
vech \left [\left\{(X_{i1}^* - \mu_X)' (X_{i1}^* - \mu_X)- \Sigma_{X^*}\right\} + \left\{(X_{i2}^* - \mu_X)^\prime (X_{i2}^* - \mu_X) - \Sigma_{X^*}\right\}  V_i \right ] \\
Z_i - \mu_Z\\
vech \left [(Z_i - \mu_Z)^\prime(Z_i - \mu_Z) - \Sigma_Z \right ]\\
vech \left [\left\{(X_{i1}^* - \mu_X)^\prime(Z_i - \mu_Z) - \Sigma_{X^\star Z}\right\} + \left\{(X_{i2}^* - \mu_X)^\prime(Z_i - \mu_Z) - \Sigma_{X^\star Z}\right\}  V_i  \right ] \\
vech \left [\left\{(X_{i1}^* - X_{i2}^*)^\prime(X_{i1}^* - X_{i2}^*) /2 - \Sigma_T\right\}  V_i  \right ] \\
vech \left [\left\{(X_{i1}^* - X_{i2}^*)^\prime(Y_{i1}^* - Y_{i2}^*)/2 - \Sigma_{T\,\tilde{T}}\right\}  V_i  \right ]
  \end{cases}
\end{equation*}
\normalsize
\noindent and the $\hat{X}_i$'s and $\hat{Y}_i$'s in the first two equations of $\psi_i(\theta)$ are functions of $\theta$, as follows. Subjects will have a different estimate of $\hat{X}$ and $\hat{Y}$, depending on whether they are in the reliability subset. We also take advantage of the parameter equalities implied by the measurement error assumptions, as described above. 

For subjects with only 1 measure, 
\begin{align*}
\hat{X}_i &= \mu_X + \begin{bmatrix}\Sigma_X & \Sigma_{ZX} \end{bmatrix}\begin{bmatrix}\Sigma_{X^*} & \Sigma_{X^\star Z} \\ \Sigma_{ZX^\star} & \Sigma_Z \end{bmatrix}^{-1} \begin{bmatrix}X_i^* - \mu_X \\ Z_i - \mu_Z\end{bmatrix},  \\
c_i^* &= E(\tilde{T} | X_i^*,Z) = \begin{bmatrix} \Sigma_{\tilde T {T}} & 0 \end{bmatrix} \begin{bmatrix}\Sigma_{X^*} & \Sigma_{X^\star Z} \\ \Sigma_{ZX^\star} & \Sigma_Z \end{bmatrix}^{-1} \begin{bmatrix}X_i^* - \mu_X \\ Z_i - \mu_Z\end{bmatrix},  \mbox{ and} \\
\hat{Y}_i &= Y_i^* - c_i^*.
\end{align*}

\noindent For subjects with reliability measure, 

\begin{align*}
{X}_i^{R} &= \mu_X + \begin{bmatrix}\Sigma_X & \Sigma_X & \Sigma_{ZX}\end{bmatrix} \begin{bmatrix}\Sigma_{X^*} & \Sigma_{X} & \Sigma_{X Z} \\ \Sigma_X & \Sigma_{X^*} & \Sigma_{XZ} \\ \Sigma_{ZX} & \Sigma_{ZX} & \Sigma_Z \end{bmatrix}^{-1} \begin{bmatrix}X_{i1}^* - \mu_X \\ X_{i2}^* - \mu_X \\ Z_i - \mu_Z\end{bmatrix},\\
c_{ij}^* &= E(\tilde{T}_{ij} | X_{i1}^*, X_{i2}^*,Z) = \begin{bmatrix} \Sigma_{\tilde T_{ij}X_{i1}^\star } & \Sigma_{\tilde T_{ij}X_{i2}^\star }  & 0\end{bmatrix}\begin{bmatrix}\Sigma_{X^*} & \Sigma_{X} & \Sigma_{X Z} \\ \Sigma_X & \Sigma_{X^*} & \Sigma_{XZ} \\ \Sigma_{ZX} & \Sigma_{ZX} & \Sigma_Z \end{bmatrix}^{-1} \begin{bmatrix}X_{i1}^* - \mu_X \\ X_{i2}^* - \mu_X \\ Z_i - \mu_Z\end{bmatrix}, \\
\hat{Y}_{ij}^{R} &= Y_{ij}^* - c_{ij}^*, \mbox{ for } j=1,2.
\end{align*}


\noindent Note, by assumption, one has $\mbox{Cov}(\tilde T_{ij}, \tilde X^\star_{ij'}) = \mbox{Cov}(\tilde T_{ij}, T_{ij'})$, which equals 0 for $j \ne j'$.  

For \textbf{case 2}, the validation subset, the parameter vector also includes the additional parameters $\Sigma_{TZ}$, $\Sigma_{\tilde{T}Z}$ and a different M-estimation vector that allows for a more general covariance structure between $(X^\star,Z)$ and $(T,\tilde{T})$. In this case, $T$ and $\tilde T$ are allowed to be correlated with $Z$. We also use the equality $\Sigma_{TZ} = \Sigma_{X^\star Z} -\Sigma_{XZ}$. In this case, we define $\theta = (\beta_0, \beta_x, \beta_z, \mu_{X^\star}, \Sigma_{X^*}, \allowbreak \mu_Z, \allowbreak \Sigma_Z , \allowbreak \Sigma_{X^\star Z}, \allowbreak \Sigma_T, \allowbreak \Sigma_{T\, \tilde{T}}, \allowbreak \Sigma_{X Z},\Sigma_{\tilde{T}Z})$ and 
\begin{equation*}
  \psi_i(\theta)=\begin{cases}
    \hat{Y}_{i} - \beta_0 - \beta_x \hat{X}_{i} - \beta_z Z_i \\
    (\hat{Y}_{i} - \beta_0 - \beta_x \hat{X}_{i} - \beta_z Z_i)\hat{X}_{i} \\
    (\hat{Y}_{i} - \beta_0 - \beta_x \hat{X}_{i} - \beta_z Z_i)Z_i\\
X_{i}^\star  (1-V_i) + X_{i}   V_i - \mu_{X^\star} \\
vech \left [ (X_{i}^* - \mu_X)^\prime (X_{i}^* - \mu_X) - \Sigma_{X^*}  \right ] \\
Z_i - \mu_Z\\
vech \left [ (Z_i - \mu_Z)^\prime (Z_i - \mu_Z) - \Sigma_Z  \right ]\\
vech \left [ (X_{i}^* - \mu_X)^\prime(Z_i - \mu_Z) - \Sigma_{X^\star Z}  \right ]\\
vech \left [ \left\{(X_{i}^* - X_{i})^\prime (X_{i}^* - X_{i}) - \Sigma_T\right\}  V_i  \right ]\\
vech \left [ \left\{(X_{i}^* - X_{i})^\prime (Y_{i}^* - Y_{i}) - \Sigma_{T\,\tilde{T}}\right\}  V_i  \right ] \\
vech \left [ \{(X_{i} - \mu_X)^\prime (Z_i - \mu_Z) - \Sigma_{XZ}\}   V_i  \right ]\\
vech \left [ \left\{(Y_{i}^* - Y_{i}) (Z_{i} - \mu_Z)- \Sigma_{\tilde{T}Z}\right\}  V_i  \right ]. 
  \end{cases} 
\end{equation*}

\noindent Here, $\hat Y_i = Y^\star_i - \hat c_i$, where $\hat c_i=\hat E[\tilde T|X^\star,Z]$ is provided by equation (\ref{estttilde}) and $\hat X$ is defined as in equation (\ref{estX}).

\noindent For \textbf{case 3}, the parameters needed to estimate $\hat{X}$ and $c^\star$ can be estimated by standard linear regression on the biomarker subset and standard errors for the proposed method in the text (Sections 3 and 4) were calculated  using the bootstrap. We provide stacked estimating equations for the sandwich variance estimate, whose performance was compared to the bootstrap variance estimator in Supplementary Table S8.  

We estimate $\hat{X}_i$ with this regression on the biomarker subset $\hat{X} = E[X_B|X^\star, Z] = \gamma_0 + \gamma_x X^\star + \gamma_z Z. $
\noindent We estimate $\hat{Y_i} = Y_i^\star - \hat c_i^\star$, where  $c^\star$ is obtained from the regression $c^\star  = E[Y^\star-Y_B |X^\star, Z] = \alpha_0 + \alpha_x X^\star + \alpha_z Z. $ We define the parameter vector $\theta = (\beta_0, \beta_x, \beta_z, \gamma_0, \gamma_x,\gamma_z,\alpha_0,\alpha_x, \alpha_z)$  and 

\begin{equation*}
  \psi_i(\theta)=\begin{cases}
    \hat{Y}_{i} - \beta_0 - \beta_x \hat{X}_{i} - \beta_z Z_i \\
    (\hat{Y}_{i} - \beta_0 - \beta_x \hat{X}_{i} - \beta_z Z_i)\hat{X}_{i} \\
    (\hat{Y}_{i} - \beta_0 - \beta_x \hat{X}_{i} - \beta_z Z_i)Z_i\\
   V_i ( X_{Bi} - \gamma_0 - \gamma_x X^{\star}_{i} - \gamma_Z Z_i) \\
    V_i (X_{Bi} - \gamma_0 - \gamma_x X^{\star}_{i} - \gamma_Z Z_i)X^{\star}_{i} \\
    V_i (X_{Bi} - \gamma_0 - \gamma_x X^{\star}_{i} - \gamma_Z Z_i)Z_i\\
       V_i ( Y_i^\star - Y_{Bi} - \alpha_0 - \alpha_x X^{\star}_{i} - \alpha_Z Z_i) \\
    V_i (Y_i^\star - Y_{Bi} - \alpha_0 - \alpha_x X^{\star}_{i} - \alpha_Z Z_i)X^{\star}_{i} \\
    V_i (Y_i^\star - Y_{Bi} - \alpha_0 - \alpha_x X^{\star}_{i} - \alpha_Z Z_i)Z_i.
  \end{cases} 
\end{equation*}


\section{ Supplementary Materials Web Appendix}
\subsection{Supplementary tables for simulation study}

The first set of supplementary tables are variations of the simulation results Table 3 described in Section 4 of the main manuscript. Table S1 repeats the scenarios in Table 3 with smaller measurement error variance parameters $\sigma_T= \sigma_{\tilde T}=0.5$.  Table S2 repeats a similar set of scenarios, changing only the size of the cohort to N=1000 and the choices for the size of the reliability subset to be n=50, 100, 200, 500, 1000.  Table S3 is the same as Table 3, except it considers negative correlation for the error terms $T$ in $X^\star$ and $\tilde T$ in $Y^\star$, namely allows $\rho_{T \tilde T} = -0.25, -0.5$. Table S4 is the same as upper half of Table 3, except it considers a larger $\beta_x=5$ and $\beta_z=-5$. Table S5 is the same as Table 3, only non-Gaussian distributions are used to simulate the error in $(X^\star, Y^\star)$. Tables S6a and S6b compares proposed regression calibration method to the moment correction method of Shepherd and Yu (2011) for validation data; Table S6a shows results for simple linear regression and Table S6b shows results for an ANCOVA model.   Simulation results in Table S7 compare the estimates of regression calibration to those from an extension of moment correction method of Shepherd and Yu (2011) to reliability data, with parameters as in Table 3. Table S8 compares performance of proposed method using a bootstrapped versus sandwich estimator for the standard error in the WHI simulations of Section 5. Figure S1 shows the error distributions considered for Table 3 (normal) and Table S5 (mixture of non-normals and log-normal). 
\setcounter{table}{0}

\begin{table}[ht]
\footnotesize
  \renewcommand{\thetable}{S\arabic{table}}
\caption{Across 1000 simulations, the percent (\%) bias, empirical standard
error (SE), average estimator for standard error (ASE), mean squared error
(MSE) and estimated 95\% coverage probability (CP) for $\beta$. Results are provided for linear regression using the true data $(X,Y)$ (TRUE Method), the error prone data $(X^\star,Y^\star)$ (NAIVE Method), and the regression calibration estimates $(\hat{X},\hat{Y})$ (Method denoted by size of reliability subset $n=25,50,100,200,400$). Data are generated according to Scenario 2  in Section 4, with $\beta_x=1$, $\beta_z=-1$ and $\sigma_T=\sigma_{\tilde T}=0.5$.}
\vspace{-.1in}
\begin{center}
\begin{tabular}{lcccccccccccc}
\hline
\hline
&  & & \multicolumn{5}{c}{$\beta_x$} & \multicolumn{5}{c}{$\beta_z$}  \\
\cline{4-8}
\cline{9-13}
  $\rho_{xz}$ & $\rho_{T \tilde T}$ & Method & \% Bias & SE & ASE & MSE & CP & \% Bias & SE & ASE & MSE & CP \\ 
  \hline
  0.5 & 0.5 & TRUE & -1.05 & 0.29 & 0.29 & 0.08 & 0.955 & 0.39 & 0.29 & 0.29 & 0.08 & 0.949 \\ 
   &  & NAIVE & -13.61 & 0.25 & 0.25 & 0.08 & 0.919 & 6.67 & 0.28 & 0.28 & 0.08 & 0.943 \\ 
   &  & 25 & -1.51 & 0.35 & 0.35 & 0.12 & 0.946 & 0.44 & 0.31 & 0.31 & 0.10 & 0.95 \\ 
   &  & 50 & -1.22 & 0.35 & 0.35 & 0.12 & 0.942 & 0.25 & 0.31 & 0.31 & 0.10 & 0.944 \\ 
   &  & 100 & -0.70 & 0.35 & 0.35 & 0.13 & 0.943 & 0.00 & 0.32 & 0.32 & 0.10 & 0.945 \\ 
   &  & 200 & -0.79 & 0.34 & 0.34 & 0.12 & 0.933 & 0.41 & 0.32 & 0.31 & 0.10 & 0.943 \\ 
   &  & 400 & -0.91 & 0.31 & 0.31 & 0.10 & 0.95 & 0.28 & 0.30 & 0.29 & 0.09 & 0.945 \\  
   & 0.25 & TRUE & -1.05 & 0.29 & 0.29 & 0.08 & 0.955 & 0.39 & 0.29 & 0.29 & 0.08 & 0.949 \\ 
   &  & NAIVE & -19.8 & 0.25 & 0.25 & 0.10 & 0.882 & 9.77 & 0.28 & 0.28 & 0.09 & 0.938 \\ 
   &  & 25 & -1.29 & 0.36 & 0.36 & 0.13 & 0.948 & 0.33 & 0.31 & 0.31 & 0.10 & 0.953 \\ 
   &  & 50 & -0.98 & 0.35 & 0.35 & 0.13 & 0.946 & 0.13 & 0.32 & 0.31 & 0.10 & 0.947 \\ 
   &  & 100 & -0.56 & 0.36 & 0.35 & 0.13 & 0.938 & -0.06 & 0.32 & 0.32 & 0.11 & 0.942 \\ 
   &  & 200 & -0.72 & 0.34 & 0.34 & 0.12 & 0.934 & 0.38 & 0.32 & 0.31 & 0.10 & 0.941 \\ 
   &  & 400 & -0.90 & 0.31 & 0.31 & 0.10 & 0.948 & 0.28 & 0.30 & 0.30 & 0.09 & 0.947 \\  
   & 0 & TRUE & -1.05 & 0.29 & 0.29 & 0.08 & 0.955 & 0.39 & 0.29 & 0.29 & 0.08 & 0.949 \\ 
   &  & NAIVE & -25.98 & 0.25 & 0.25 & 0.13 & 0.822 & 12.86 & 0.28 & 0.28 & 0.10 & 0.93 \\ 
   &  & 25 & -1.04 & 0.36 & 0.36 & 0.13 & 0.951 & 0.21 & 0.31 & 0.31 & 0.10 & 0.952 \\ 
   &  & 50 & -0.73 & 0.36 & 0.35 & 0.13 & 0.948 & 0.01 & 0.32 & 0.32 & 0.10 & 0.948 \\ 
   &  & 100 & -0.43 & 0.36 & 0.35 & 0.13 & 0.944 & -0.12 & 0.33 & 0.32 & 0.11 & 0.945 \\ 
   &  & 200 & -0.65 & 0.34 & 0.34 & 0.12 & 0.938 & 0.35 & 0.32 & 0.31 & 0.10 & 0.941 \\ 
   &  & 400 & -0.88 & 0.31 & 0.31 & 0.10 & 0.948 & 0.27 & 0.30 & 0.30 & 0.09 & 0.947 \\ 
  0 & 0.5 & TRUE & -0.91 & 0.24 & 0.25 & 0.06 & 0.955 & 0.10 & 0.25 & 0.25 & 0.06 & 0.947 \\ 
   &  & NAIVE & -10.99 & 0.22 & 0.23 & 0.06 & 0.923 & 0.11 & 0.25 & 0.25 & 0.06 & 0.95 \\ 
   &  & 25 & -1.35 & 0.29 & 0.29 & 0.08 & 0.946 & -0.04 & 0.26 & 0.26 & 0.07 & 0.947 \\ 
   &  & 50 & -1.15 & 0.29 & 0.29 & 0.08 & 0.947 & -0.07 & 0.26 & 0.26 & 0.07 & 0.94 \\ 
   &  & 100 & -0.72 & 0.29 & 0.29 & 0.09 & 0.949 & -0.13 & 0.27 & 0.26 & 0.07 & 0.941 \\ 
   &  & 200 & -0.64 & 0.29 & 0.29 & 0.08 & 0.942 & 0.19 & 0.26 & 0.26 & 0.07 & 0.949 \\ 
   &  & 400 & -0.81 & 0.26 & 0.26 & 0.07 & 0.951 & 0.06 & 0.25 & 0.25 & 0.06 & 0.946 \\ 
   & 0.25 & TRUE & -0.91 & 0.24 & 0.25 & 0.06 & 0.955 & 0.10 & 0.25 & 0.25 & 0.06 & 0.947 \\ 
   &  & NAIVE & -15.93 & 0.22 & 0.23 & 0.08 & 0.889 & 0.11 & 0.25 & 0.25 & 0.06 & 0.948 \\ 
   &  & 25 & -1.22 & 0.29 & 0.29 & 0.09 & 0.946 & -0.03 & 0.26 & 0.26 & 0.07 & 0.948 \\ 
   &  & 50 & -0.99 & 0.29 & 0.29 & 0.09 & 0.946 & -0.07 & 0.26 & 0.26 & 0.07 & 0.94 \\ 
   &  & 100 & -0.62 & 0.29 & 0.29 & 0.09 & 0.947 & -0.13 & 0.27 & 0.26 & 0.07 & 0.94 \\ 
   &  & 200 & -0.59 & 0.29 & 0.29 & 0.08 & 0.942 & 0.19 & 0.26 & 0.26 & 0.07 & 0.947 \\ 
   &  & 400 & -0.8 & 0.26 & 0.27 & 0.07 & 0.95 & 0.06 & 0.25 & 0.25 & 0.06 & 0.947 \\ 
   & 0 & TRUE & -0.91 & 0.24 & 0.25 & 0.06 & 0.955 & 0.10 & 0.25 & 0.25 & 0.06 & 0.947 \\ 
   &  & NAIVE & -20.88 & 0.22 & 0.23 & 0.09 & 0.85 & 0.10 & 0.25 & 0.25 & 0.06 & 0.948 \\ 
   &  & 25 & -1.07 & 0.30 & 0.30 & 0.09 & 0.944 & -0.02 & 0.26 & 0.26 & 0.07 & 0.95 \\ 
   &  & 50 & -0.82 & 0.29 & 0.30 & 0.09 & 0.946 & -0.07 & 0.26 & 0.26 & 0.07 & 0.941 \\ 
   &  & 100 & -0.53 & 0.30 & 0.29 & 0.09 & 0.944 & -0.13 & 0.27 & 0.26 & 0.07 & 0.94 \\ 
   &  & 200 & -0.54 & 0.29 & 0.29 & 0.08 & 0.939 & 0.19 & 0.26 & 0.26 & 0.07 & 0.946 \\ 
   &  & 400 & -0.79 & 0.26 & 0.27 & 0.07 & 0.948 & 0.06 & 0.25 & 0.25 & 0.06 & 0.947 \\ 
   \hline
    \end{tabular}
  \end{center}
\end{table}
\normalsize

\clearpage

\begin{table}[ht]
\footnotesize
  \renewcommand{\thetable}{S\arabic{table}}
\caption{Across 1000 simulations, the percent (\%) bias, empirical standard
error (SE), average estimator for standard error (ASE), mean squared error
(MSE) and estimated 95\% coverage probability (CP) for $\beta$. Results are provided for linear regression using the true data $(X,Y)$ (TRUE Method), the error prone data $(X^\star,Y^\star)$ (NAIVE Method), and the regression calibration estimates $(\hat{X},\hat{Y})$ (Method denoted by size of reliability subset $n=50,100,200,500,1000$). Data are generated according to Scenario 2  in Section 4, with $\beta_x=1$, $\beta_z=-1$ and $\sigma_T=\sigma_{\tilde T}=1$, and N=1000.}
\begin{center}
\begin{tabular}{lcccccccccccc}
\hline
\hline
&  & & \multicolumn{5}{c}{$\beta_x$} & \multicolumn{5}{c}{$\beta_z$}  \\
\cline{4-8}
\cline{9-13}
  $\rho_{xz}$ & $\rho_{T \tilde T}$ & Method & \% Bias & SE & ASE & MSE & CP & \% Bias & SE & ASE & MSE & CP \\ 
  \hline
   0.5 & 0.5 & TRUE & 0.61 & 0.19 & 0.18 & 0.03 & 0.948 & 0.52 & 0.18 & 0.18 & 0.03 & 0.955 \\ 
   &  & NAIVE & -27.85 & 0.12 & 0.12 & 0.09 & 0.358 & 14.7 & 0.16 & 0.17 & 0.05 & 0.875 \\ 
   &  & 50 & 5.94 & 0.45 & 0.44 & 0.21 & 0.975 & -2.18 & 0.27 & 0.28 & 0.08 & 0.977 \\ 
   &  & 100 & 3.46 & 0.35 & 0.34 & 0.12 & 0.965 & -0.81 & 0.23 & 0.24 & 0.05 & 0.966 \\ 
   &  & 200 & 2.33 & 0.30 & 0.31 & 0.09 & 0.955 & -0.12 & 0.22 & 0.23 & 0.05 & 0.958 \\ 
   &  & 500 & 1.13 & 0.26 & 0.28 & 0.07 & 0.95 & -0.09 & 0.21 & 0.22 & 0.04 & 0.956 \\ 
   &  & 1000 & 1.47 & 0.23 & 0.24 & 0.06 & 0.956 & 0.00 & 0.19 & 0.20 & 0.04 & 0.96 \\  
   
   & 0.25 & TRUE & 0.61 & 0.19 & 0.18 & 0.03 & 0.948 & 0.52 & 0.18 & 0.18 & 0.03 & 0.955 \\ 
   &  & NAIVE & -42.11 & 0.12 & 0.12 & 0.19 & 0.066 & 21.82 & 0.16 & 0.17 & 0.07 & 0.762 \\ 
   &  & 50 & 7.92 & 0.50 & 0.49 & 0.26 & 0.953 & -3.17 & 0.30 & 0.31 & 0.09 & 0.97 \\ 
   &  & 100 & 4.46 & 0.38 & 0.37 & 0.15 & 0.958 & -1.32 & 0.25 & 0.25 & 0.06 & 0.966 \\ 
   &  & 200 & 2.93 & 0.31 & 0.32 & 0.10 & 0.958 & -0.42 & 0.22 & 0.23 & 0.05 & 0.958 \\ 
   &  & 500 & 1.3 & 0.27 & 0.28 & 0.07 & 0.952 & -0.19 & 0.21 & 0.22 & 0.04 & 0.956 \\ 
   &  & 1000 & 1.54 & 0.24 & 0.24 & 0.06 & 0.954 & -0.04 & 0.19 & 0.20 & 0.04 & 0.961 \\  
   
   & 0 & TRUE & 0.61 & 0.19 & 0.18 & 0.03 & 0.948 & 0.52 & 0.18 & 0.18 & 0.03 & 0.955 \\ 
   &  & NAIVE & -56.38 & 0.12 & 0.12 & 0.33 & 0.006 & 28.94 & 0.16 & 0.17 & 0.11 & 0.625 \\ 
   &  & 50 & 10.17 & 0.57 & 0.55 & 0.33 & 0.936 & -4.28 & 0.33 & 0.34 & 0.11 & 0.961 \\ 
   &  & 100 & 5.47 & 0.41 & 0.40 & 0.17 & 0.951 & -1.82 & 0.26 & 0.27 & 0.07 & 0.968 \\ 
   &  & 200 & 3.52 & 0.33 & 0.33 & 0.11 & 0.958 & -0.72 & 0.23 & 0.24 & 0.05 & 0.959 \\ 
   &  & 500 & 1.48 & 0.27 & 0.28 & 0.08 & 0.954 & -0.29 & 0.21 & 0.22 & 0.05 & 0.958 \\ 
   &  & 1000 & 1.61 & 0.24 & 0.25 & 0.06 & 0.949 & -0.07 & 0.19 & 0.20 & 0.04 & 0.962 \\ 
   
  0 & 0.5 & TRUE & 0.72 & 0.16 & 0.16 & 0.03 & 0.95 & 0.66 & 0.15 & 0.16 & 0.02 & 0.958 \\ 
   &  & NAIVE & -24.25 & 0.11 & 0.11 & 0.07 & 0.421 & 0.62 & 0.15 & 0.16 & 0.02 & 0.961 \\ 
   &  & 50 & 3.29 & 0.30 & 0.30 & 0.09 & 0.966 & 0.68 & 0.16 & 0.16 & 0.03 & 0.957 \\ 
   &  & 100 & 2.44 & 0.27 & 0.26 & 0.07 & 0.955 & 0.76 & 0.16 & 0.17 & 0.03 & 0.955 \\ 
   &  & 200 & 1.93 & 0.24 & 0.25 & 0.06 & 0.953 & 0.93 & 0.16 & 0.17 & 0.03 & 0.958 \\ 
   &  & 500 & 0.95 & 0.21 & 0.22 & 0.05 & 0.952 & 0.40 & 0.16 & 0.17 & 0.03 & 0.964 \\ 
   &  & 1000 & 1.33 & 0.19 & 0.20 & 0.04 & 0.954 & 0.59 & 0.15 & 0.16 & 0.02 & 0.961 \\ 
   
   & 0.25 & TRUE & 0.72 & 0.16 & 0.16 & 0.03 & 0.95 & 0.66 & 0.15 & 0.16 & 0.02 & 0.958 \\ 
   &  & NAIVE & -36.73 & 0.11 & 0.11 & 0.15 & 0.108 & 0.62 & 0.15 & 0.16 & 0.02 & 0.961 \\ 
   &  & 50 & 4.31 & 0.34 & 0.33 & 0.12 & 0.957 & 0.68 & 0.16 & 0.17 & 0.03 & 0.957 \\ 
   &  & 100 & 2.97 & 0.29 & 0.28 & 0.08 & 0.958 & 0.75 & 0.16 & 0.17 & 0.03 & 0.956 \\ 
   &  & 200 & 2.3 & 0.25 & 0.25 & 0.06 & 0.955 & 0.93 & 0.16 & 0.17 & 0.03 & 0.959 \\ 
   &  & 500 & 1.04 & 0.22 & 0.23 & 0.05 & 0.953 & 0.39 & 0.16 & 0.17 & 0.03 & 0.964 \\ 
   &  & 1000 & 1.37 & 0.20 & 0.20 & 0.04 & 0.952 & 0.59 & 0.15 & 0.16 & 0.02 & 0.961 \\ 
   
   & 0 & TRUE & 0.72 & 0.16 & 0.16 & 0.03 & 0.95 & 0.66 & 0.15 & 0.16 & 0.02 & 0.958 \\ 
   &  & NAIVE & -49.22 & 0.12 & 0.12 & 0.26 & 0.01 & 0.61 & 0.16 & 0.16 & 0.02 & 0.962 \\ 
   &  & 50 & 5.38 & 0.38 & 0.37 & 0.15 & 0.946 & 0.68 & 0.16 & 0.17 & 0.03 & 0.959 \\ 
   &  & 100 & 3.5 & 0.31 & 0.30 & 0.10 & 0.958 & 0.75 & 0.16 & 0.17 & 0.03 & 0.956 \\ 
   &  & 200 & 2.66 & 0.26 & 0.26 & 0.07 & 0.959 & 0.93 & 0.16 & 0.17 & 0.03 & 0.962 \\ 
   &  & 500 & 1.14 & 0.22 & 0.23 & 0.05 & 0.955 & 0.38 & 0.16 & 0.17 & 0.03 & 0.963 \\ 
   &  & 1000 & 1.41 & 0.20 & 0.20 & 0.04 & 0.952 & 0.59 & 0.15 & 0.16 & 0.02 & 0.96 \\  \hline

    \end{tabular}

  \end{center}
\end{table}
\normalsize
\clearpage

\begin{table}[ht]
\footnotesize
  \renewcommand{\thetable}{S\arabic{table}}
\caption{Across 1000 simulations, the percent (\%) bias, empirical standard
error (SE), average estimator for standard error (ASE), mean squared error
(MSE) and estimated 95\% coverage probability (CP) for $\beta$. Results are provided for linear regression using the true data $(X,Y)$ (TRUE Method), the error prone data $(X^\star,Y^\star)$ (NAIVE Method), and the regression calibration estimates  $(\hat{X},\hat{Y})$ (Method denoted by size of reliability subset $n=25,50,100,200,400$). Data are generated according to Scenario 2 in Section 4, with $\beta_x=1$, $\beta_z=-1$ and $\sigma_T=\sigma_{\tilde T}=1$, N=400. Negative correlations are considered for $\rho_{T \tilde T}$.} 
\begin{center}
\begin{tabular}{lcccccccccccc}
\hline
\hline
&  & & \multicolumn{5}{c}{$\beta_x$} & \multicolumn{5}{c}{$\beta_z$}  \\
\cline{4-8}
\cline{9-13}
  $\rho_{xz}$ & $\rho_{T \tilde T}$ & Method & \% Bias & SE & ASE & MSE & CP & \% Bias & SE & ASE & MSE & CP \\ 
  \hline
0.5 & -0.5 & TRUE & -1.05 & 0.29 & 0.29 & 0.08 & 0.955 & 0.39 & 0.29 & 0.29 & 0.08 & 0.949 \\ 
   &  & NAIVE & -86.12 & 0.20 & 0.20 & 0.78 & 0.012 & 43.04 & 0.27 & 0.28 & 0.26 & 0.665 \\ 
   &  & 25 & 57.1 & 11.26 & 66.83 & 127.18 & 0.93 & -28.52 & 5.89 & 35.33 & 34.75 & 0.959 \\ 
   &  & 50 & 13.18 & 0.83 & 0.83 & 0.71 & 0.942 & -6.75 & 0.50 & 0.52 & 0.25 & 0.965 \\ 
   &  & 100 & 5.58 & 0.59 & 0.57 & 0.35 & 0.958 & -3.01 & 0.41 & 0.40 & 0.17 & 0.951 \\ 
   &  & 200 & 1.83 & 0.48 & 0.48 & 0.23 & 0.958 & -0.84 & 0.36 & 0.36 & 0.13 & 0.943 \\ 
   &  & 400 & 0.30 & 0.40 & 0.40 & 0.16 & 0.953 & -0.32 & 0.32 & 0.33 & 0.10 & 0.955 \\ \\
   
   & -0.25 & TRUE & -1.05 & 0.29 & 0.29 & 0.08 & 0.955 & 0.39 & 0.29 & 0.29 & 0.08 & 0.949 \\ 
   &  & NAIVE & -71.96 & 0.20 & 0.20 & 0.56 & 0.051 & 35.95 & 0.27 & 0.28 & 0.20 & 0.738 \\ 
   &  & 25 & 9.63 & 4.54 & 11.85 & 20.62 & 0.939 & -4.30 & 2.38 & 6.26 & 5.67 & 0.959 \\ 
   &  & 50 & 10.21 & 0.75 & 0.75 & 0.57 & 0.952 & -5.24 & 0.46 & 0.48 & 0.21 & 0.961 \\ 
   &  & 100 & 4.55 & 0.57 & 0.55 & 0.32 & 0.956 & -2.51 & 0.40 & 0.39 & 0.16 & 0.946 \\ 
   &  & 200 & 1.38 & 0.47 & 0.47 & 0.22 & 0.951 & -0.63 & 0.36 & 0.36 & 0.13 & 0.943 \\ 
   &  & 400 & 0.13 & 0.39 & 0.39 & 0.15 & 0.949 & -0.25 & 0.32 & 0.32 & 0.10 & 0.956 \\ \\
   
  0 & -0.5 & TRUE & -0.91 & 0.24 & 0.25 & 0.06 & 0.955 & 0.10 & 0.25 & 0.25 & 0.06 & 0.947 \\ 
   &  & NAIVE & -75.38 & 0.18 & 0.18 & 0.6 & 0.026 & 0.11 & 0.26 & 0.26 & 0.07 & 0.95 \\ 
   &  & 25 & 28.26 & 2.54 & 2.89 & 6.52 & 0.936 & 1.18 & 0.33 & 0.40 & 0.11 & 0.949 \\ 
   &  & 50 & 10.06 & 1.27 & 1.00 & 1.63 & 0.947 & -0.04 & 0.28 & 0.29 & 0.08 & 0.95 \\ 
   &  & 100 & 2.91 & 0.44 & 0.43 & 0.19 & 0.951 & -0.06 & 0.28 & 0.28 & 0.08 & 0.951 \\ 
   &  & 200 & 1.04 & 0.38 & 0.38 & 0.14 & 0.951 & 0.16 & 0.27 & 0.27 & 0.07 & 0.944 \\ 
   &  & 400 & -0.01 & 0.32 & 0.32 & 0.10 & 0.953 & 0.01 & 0.25 & 0.26 & 0.06 & 0.95 \\ \\
   
   & -0.25 & TRUE & -0.91 & 0.24 & 0.25 & 0.06 & 0.955 & 0.10 & 0.25 & 0.25 & 0.06 & 0.947 \\ 
   &  & NAIVE & -62.99 & 0.18 & 0.18 & 0.43 & 0.07 & 0.13 & 0.26 & 0.26 & 0.07 & 0.948 \\ 
   &  & 25 & 11.53 & 5.23 & 12.07 & 27.38 & 0.937 & 0.75 & 0.41 & 0.85 & 0.17 & 0.948 \\ 
   &  & 50 & -0.32 & 1.73 & 1.58 & 3.00 & 0.954 & 0.28 & 0.28 & 0.31 & 0.08 & 0.951 \\ 
   &  & 100 & 2.29 & 0.43 & 0.42 & 0.18 & 0.95 & -0.06 & 0.27 & 0.27 & 0.08 & 0.947 \\ 
   &  & 200 & 0.76 & 0.37 & 0.37 & 0.14 & 0.949 & 0.16 & 0.27 & 0.27 & 0.07 & 0.944 \\ 
   &  & 400 & -0.12 & 0.32 & 0.32 & 0.10 & 0.95 & 0.01 & 0.25 & 0.26 & 0.06 & 0.951 \\ 

    \hline
    \end{tabular}

  \end{center}
\end{table}
\normalsize

\clearpage

\begin{table}[ht]
\centering
\footnotesize
  \renewcommand{\thetable}{S\arabic{table}}
\caption{Across 1000 simulations, the percent bias (\%), empirical standard
error (SE), average estimated standard error (ASE), mean squared error
(MSE) and estimated 95\% coverage probability (CP) for $\beta$. Results are provided for linear regression using the true data $(X,Y)$ (TRUE Method), the error prone data $(X^\star,Y^\star)$ (NAIVE Method), and the regression calibration estimates  $(\hat{X},\hat{Y})$ (Method denoted by size of reliability subset $n=25,50,100,200,400$). Data are generated according to Scenario 2  in Section 4, with $\beta_x=5$, $\beta_z=-5$ and $\sigma_T=\sigma_{\tilde T}=1$, N=400. } 
\begin{tabular}{lccccccccccccccccc}
  \hline
&&\multicolumn{5}{c}{$\beta_x$} & \multicolumn{5}{c}{$\beta_z$}\\
\cline{3-7}
\cline{8-13}
 $\rho_{xz}$ &$\rho_{T \tilde T}$ & Method & \% Bias & SE & ASE & MSE & CP & \% Bias & SE & ASE & MSE & CP \\
  \hline
0.5 & 0.5& True & -0.21 & 0.29 & 0.29 & 0.08 & 0.95 & 0.08 & 0.29 & 0.29 & 0.08 & 0.95 \\ 
  &  & NAIVE & -51.48 & 0.22 & 0.22 & 6.68 & 0.00 & 25.70 & 0.31 & 0.32 & 1.75 & 0.02 \\ 
   & & 25 & 10.27 & 16.00 & 37.51 & 256.20 & 0.85 & -5.16 & 7.35 & 17.38 & 54.14 & 0.90 \\ 
    && 50 & 7.58 & 2.77 & 2.18 & 7.81 & 0.90 & -3.67 & 1.38 & 1.17 & 1.93 & 0.94 \\ 
   & & 100 & 2.99 & 1.12 & 1.03 & 1.27 & 0.93 & -1.51 & 0.65 & 0.63 & 0.43 & 0.96 \\ 
  &  & 200 & 1.42 & 0.71 & 0.70 & 0.50 & 0.95 & -0.66 & 0.48 & 0.48 & 0.23 & 0.96 \\ 
   & & 400 & 0.63 & 0.52 & 0.52 & 0.28 & 0.95 & -0.31 & 0.39 & 0.40 & 0.15 & 0.96 \\ 
  &  0.25 & True & -0.21 & 0.29 & 0.29 & 0.08 & 0.95 & 0.08 & 0.29 & 0.29 & 0.08 & 0.95 \\ 
   & & NAIVE & -54.33 & 0.23 & 0.23 & 7.43 & 0.00 & 27.13 & 0.32 & 0.32 & 1.94 & 0.01 \\ 
   & & 25 & 20.37 & 18.21 & 58.59 & 332.59 & 0.86 & -9.56 & 7.90 & 25.44 & 62.61 & 0.90 \\ 
   & & 50 & 7.58 & 2.19 & 1.93 & 4.96 & 0.90 & -3.70 & 1.12 & 1.05 & 1.28 & 0.93 \\ 
   & & 100 & 3.36 & 1.20 & 1.08 & 1.47 & 0.94 & -1.69 & 0.69 & 0.66 & 0.49 & 0.95 \\ 
   & & 200 & 1.58 & 0.73 & 0.73 & 0.55 & 0.95 & -0.75 & 0.49 & 0.50 & 0.25 & 0.96 \\ 
   & & 400 & 0.69 & 0.54 & 0.54 & 0.29 & 0.95 & -0.34 & 0.40 & 0.41 & 0.16 & 0.95 \\ 
   & 0 & True & -0.21 & 0.29 & 0.29 & 0.08 & 0.95 & 0.08 & 0.29 & 0.29 & 0.08 & 0.95 \\ 
  &  & NAIVE & -57.17 & 0.23 & 0.23 & 8.22 & 0.00 & 28.55 & 0.32 & 0.33 & 2.14 & 0.01 \\ 
  &  & 25 & 23.00 & 15.79 & 31.92 & 250.64 & 0.86 & -11.39 & 8.40 & 16.93 & 70.91 & 0.90 \\ 
  &  & 50 & 8.11 & 2.14 & 1.97 & 4.74 & 0.90 & -3.99 & 1.10 & 1.07 & 1.26 & 0.93 \\ 
  &  & 100 & 3.69 & 1.25 & 1.13 & 1.60 & 0.94 & -1.85 & 0.72 & 0.68 & 0.53 & 0.95 \\ 
  &  & 200 & 1.74 & 0.76 & 0.76 & 0.59 & 0.95 & -0.83 & 0.51 & 0.51 & 0.26 & 0.96 \\ 
  &  & 400 & 0.74 & 0.55 & 0.55 & 0.31 & 0.94 & -0.38 & 0.41 & 0.42 & 0.17 & 0.96 \\  
   \hline
\end{tabular}
\end{table}

\clearpage
\begin{table}[ht]

\small
  \renewcommand{\thetable}{S\arabic{table}}
\caption{Non-Gaussian distributions are considered for the distribution of the error in $(X^\star,Y^\star)$. Across 1000 simulations, the percent (\%) bias, empirical standard error (SE), average estimator for standard error (ASE), mean squared error
(MSE) and estimated 95\% coverage probability (CP) for $\beta$. Results are provided for simple linear regression using the true $(X,Y)$ (TRUE Method), the error prone data $(X^\star,Y^\star)$ (NAIVE Method), and the regression calibration estimates  $(\hat{X},\hat{Y})$ (Method denoted by size of reliability subset $n=50,100,200,500,1000$). Data are generated according to Scenario 1 in Section 4, with $\beta_x=1$ and $\sigma_T=\sigma_{\tilde T}=1$, N=1000.} 
\begin{center}
\begin{tabular}{lccccccccccccccccc}
\hline
\hline
& \multicolumn{5}{c}{Mixture of normals} & \multicolumn{5}{c}{Log-normal}  \\
\cline{2-6}
\cline{7-12}
   $\rho_{T \tilde T}$ & Method & \% Bias & SE & ASE & MSE & CP & \% Bias & SE & ASE & MSE & CP \\ 
  \hline
0.5 &TRUE & -0.59 & 0.16 & 0.16 & 0.03 & 0.95 & 0.24 & 0.16 & 0.16 & 0.03 & 0.95 \\ 
  & NAIVE & -25.2 & 0.12 & 0.11 & 0.08 & 0.41 & -26.2 & 0.12 & 0.11 & 0.08 & 0.35 \\ 
&  50 & 2.31 & 0.33 & 0.31 & 0.11 & 0.95 & -3.03 & 1.43 & 2.97 & 2.04 & 0.94 \\ 
 & 100 & 0.68 & 0.27 & 0.26 & 0.07 & 0.95 & 40.8 & 9.30 & 151 & 86.6 & 0.95 \\ 
 & 200 & -0.08 & 0.25 & 0.24 & 0.06 & 0.95 & -0.34 & 0.47 & 0.45 & 0.22 & 0.96 \\ 
 & 500 & -0.51 & 0.23 & 0.22 & 0.05 & 0.95 & 0.41 & 0.24 & 0.23 & 0.06 & 0.95 \\ 
 & 1000 & -0.41 & 0.20 & 0.20 & 0.04 & 0.95 & -0.18 & 0.20 & 0.20 & 0.04 & 0.95 \\ 
 0 & TRUE & -0.05 & 0.16 & 0.16 & 0.03 & 0.95 & 0.24 & 0.16 & 0.16 & 0.03 & 0.95 \\ 
&    NAIVE & -50.5 & 0.12 & 0.12 & 0.27 & 0.01 & -50.2 & 0.12 & 0.11 & 0.27 & 0.01 \\ 
&    50 & 3.41 & 0.37 & 0.36 & 0.14 & 0.95 & 46.9 & 9.81 & 108 & 96.5 & 0.88 \\ 
&    100 & 0.84 & 0.29 & 0.29 & 0.08 & 0.96 & 16.8 & 1.76 & 3.08 & 3.11 & 0.94 \\ 
&    200 & -0.24 & 0.25 & 0.26 & 0.06 & 0.96 & 5.17 & 0.50 & 0.44 & 0.26 & 0.96 \\ 
&    500 & -1.21 & 0.23 & 0.23 & 0.05 & 0.96 & 1.44 & 0.25 & 0.24 & 0.06 & 0.95 \\ 
&    1000 & -0.72 & 0.20 & 0.20 & 0.04 & 0.95 & 0.15 & 0.20 & 0.20 & 0.04 & 0.95 \\ 
-0.5 &    TRUE & -0.59 & 0.16 & 0.16 & 0.03 & 0.95 & 0.24 & 0.16 & 0.16 & 0.03 & 0.95 \\ 
&    NAIVE & -74.9 & 0.12 & 0.12 & 0.58 & 0.00 & -75.8 & 0.12 & 0.12 & 0.59 & 0.00 \\ 
  &  50 & 6.43 & 0.45 & 0.40 & 0.20 & 0.94 & -99.2 & 25.0 & 380 & 627 & 0.88 \\ 
  &  100 & 2.60 & 0.33 & 0.31 & 0.11 & 0.95 & 1.68 & 2.35 & 3.69 & 5.51 & 0.92 \\ 
&    200 & 1.06 & 0.27 & 0.27 & 0.07 & 0.96 & 4.02 & 0.65 & 0.63 & 0.42 & 0.95 \\ 
&    500 & -0.24 & 0.24 & 0.23 & 0.06 & 0.95 & 1.27 & 0.26 & 0.25 & 0.07 & 0.94 \\ 
&    1000 & -0.24 & 0.21 & 0.20 & 0.04 & 0.95 & -0.18 & 0.21 & 0.21 & 0.04 & 0.95 \\ 

    \hline
    \end{tabular}
  \end{center}
\end{table}
\normalsize

\clearpage

\begin{table}[ht]
 \renewcommand{\thetable}{S\arabic{table}a}
\caption{Across 1000 simulations, the percent (\%) bias, empirical standard
error (SE) and mean squared error
(MSE) for $\beta_x$. Results are provided for linear regression using the true data $(X,Y)$ (TRUE Method), the error prone data $(X^\star,Y^\star)$ (NAIVE Method), the regression calibration estimates $(\hat{X},\hat{Y})$ and using the method of moment correction in Shepherd and Yu (2011) for a validation subset $(n=25,50,100,300$). Data generated according to simulated example of HIV viral load regressed on CD4 presented in Table 2 in Shepherd and Yu (2011), with $p=0.3, \sigma_T=50, \mbox{Cor}(T,\tilde T)=0.5$, N=1000. }
\vspace{.2 in}
\begin{tabular}{lrrrrrrrrr}
  \hline
  \hline
&\multicolumn{3}{c}{Regression Calibration} & \multicolumn{3}{c}{Method of Moments} \\
  & \%Bias & SE& MSEx$10^7$ & \%Bias & SE & MSEx$10^7$\\ 
  \hline
  TRUE & 0.02 & 0.03 & 0.99 & 0.02 & 0.03 & 0.99 \\ 
  NAIVE & -34.65 & 0.05 & 122.59 & -34.65 & 0.05 & 122.59 \\ 
  25 & 2.91 & 0.30 & 91.58 & 2.91 & 0.30 & 91.58 \\ 
  50 & 1.67 & 0.20 & 40.58 & 1.67 & 0.20 & 40.58 \\ 
  100 & 1.30 & 0.14 & 20.76 & 1.30 & 0.14 & 20.76 \\ 
  300 & 0.27 & 0.08 & 7.07 & 0.27 & 0.08 & 7.07 \\ 
   \hline
\end{tabular}
\end{table}

\begin{table}[ht]
\centering
\scriptsize
 \addtocounter{table}{-1}
 \renewcommand{\thetable}{S\arabic{table}b}
\caption{Across 1000 simulations, the percent (\%) bias, empirical standard
error (SE), and mean squared error
(MSE) for $\beta_x$ and $\beta_z$. Results are provided for linear regression using the true data $(X,Y)$ (TRUE Method), the error prone data $(X^\star,Y^\star)$ (NAIVE Method), regression calibration (RC) and using the method of moment correction (MM) in Shepherd and Yu (2011) for a validation subset $(n=25,50,100,300$). Data generated according to simulated example of HIV viral load regressed on CD4 presented in Tables 15 and 18 in Shepherd and Yu (2011), with $p=0.3, \sigma_T=20 \mbox{ or }50, \mbox{Cor}(T,\tilde T)=0.5$, N=1000.}
\vspace{.1 in}

\begin{tabular}{|lc|cccccc|cccccc|}
\hline
\hline
&&\multicolumn{6}{|c|}{$\sigma_T = 50$} & \multicolumn{6}{c|}{$\sigma_T = 20$} \\
&&\multicolumn{3}{|c}{$\beta_x$} & \multicolumn{3}{c|}{$\beta_z$}&\multicolumn{3}{c}{$\beta_x$} & \multicolumn{3}{c|}{$\beta_z$} \\
  \hline
  &  & \%Bias & SE & MSE & \%Bias & SE & MSE & \%Bias & SE & MSE & \%Bias & SE & MSE \\ 
    &  &  &  &  $\mbox{ x}10^7$ & &  &$\mbox{ x}10^3$ & &  &$\mbox{ x}10^7$ & &  &$\mbox{ x}10^3$ \\ 

  & TRUE & -0.17 & 0.04 & 1.23 & -0.04 & 0.17 & 0.30 & -0.17 & 0.04 & 1.23 & -0.04 & 0.17 & 0.30 \\ 
  & NAIVE & -37.37 & 0.05 & 142.41 & -5.63 & 0.26 & 3.83 & -11.46 & 0.05 & 15.58 & -1.73 & 0.24 & 0.89 \\ 
 MM & 25 & 8.70 & 1.89 & 3572.3 & -0.09 & 3.03 & 91.99 & 0.12 & 0.11 & 12.38 & 0.00 & 0.29 & 0.83 \\ 
  & 50 & 4.63 & 0.30 & 90.17 & -0.05 & 0.52 & 2.70 & 0.21 & 0.09 & 7.42 & 0.00 & 0.27 & 0.72 \\ 
  & 100 & 2.01 & 0.19 & 35.97 & -0.02 & 0.38 & 1.49 & 0.01 & 0.07 & 4.73 & 0.00 & 0.26 & 0.66 \\ 
  & 300 & 0.90 & 0.10 & 10.93 & -0.01 & 0.30 & 0.88 & 0.00 & 0.06 & 3.10 & 0.00 & 0.25 & 0.61 \\ 

 RC & 25 & 2.64 & 0.32 & 100.0 & -0.03 & 0.60 & 3.81 & 0.26 & 0.11 & 12.48 & 0.00 & 0.27 & 0.80 \\ 
  & 50 & 1.55 & 0.22 & 48.00 & -0.02 & 0.42 & 2.03 & 0.04 & 0.09 & 7.39 & 0.00 & 0.26 & 0.74 \\ 
  & 100 & 0.11 & 0.15 & 21.61 & 0.00 & 0.33 & 1.40 & -0.35 & 0.07 & 4.49 & 0.00 & 0.25 & 0.71 \\ 
  & 300 & -0.66 & 0.09 & 7.60 & 0.01 & 0.27 & 1.10 & -0.54 & 0.06 & 3.01 & 0.01 & 0.25 & 0.70 \\ 
   \hline
\end{tabular}
\end{table}

\clearpage

\begin{sidewaystable}[ht]
 \renewcommand{\thetable}{S\arabic{table}}
\scriptsize
\vspace{-.1 in}
\begin{center}
Table S7: Comparable to Table 3 in manuscript but adds in the Moment Correction estimator adapted for reliability data for comparison.
\begin{tabular}{lcccccccccccccccccccccc}
\hline
\hline
&  & & \multicolumn{6}{c}{$\beta_x$} & \multicolumn{6}{c}{$\beta_z$}  \\
\cline{4-8}
\cline{9-15}
& & & \multicolumn{3}{c}{Moment Correction} & \multicolumn{3}{c}{Regression Calibration} &  \multicolumn{3}{c}{Moment Correction} & \multicolumn{3}{c}{Regression Calibration  }  \\
  \hline
    $\rho_{xz}$ & $\rho_{T \tilde T}$ & Method & \% Bias & SE & MSE & \% Bias & SE & MSE  & \% Bias & SE & MSE & \% Bias & SE & MSE\\ 
    0.5 & 0.5 & TRUE & -1.05 & 0.29 & 0.08 & -1.05 & 0.29 & 0.08 & 0.39 & 0.29 & 0.08 & 0.39 & 0.29 & 0.08 \\ 
   &  & NAIVE & -29.39 & 0.19 & 0.12 & -29.39 & 0.19 & 0.12 & 14.61 & 0.27 & 0.09 & 14.61 & 0.27 & 0.09 \\ 
   &  & 25 & 5.20 & 3.19 & 10.15 & 1.81 & 3.04 & 9.25 & -2.66 & 1.42 & 2.02 & -1.19 & 1.36 & 1.85 \\ 
   &  & 50 & 2.97 & 0.61 & 0.37 & 2.90 & 0.63 & 0.4 & -1.63 & 0.40 & 0.16 & -1.59 & 0.41 & 0.17 \\ 
   &  & 100 & 0.97 & 0.49 & 0.24 & 1.11 & 0.50 & 0.25 & -0.69 & 0.37 & 0.14 & -0.84 & 0.37 & 0.14 \\ 
   &  & 200 & -0.20 & 0.43 & 0.19 & -0.01 & 0.44 & 0.20 & 0.21 & 0.35 & 0.12 & 0.02 & 0.35 & 0.12 \\ 
   &  & 400 & -0.51 & 0.38 & 0.14 & -0.41 & 0.38 & 0.15 & 0.17 & 0.31 & 0.1 & 0.01 & 0.32 & 0.10 \\ 
   & 0.25 & TRUE & -1.05 & 0.29 & 0.08 & -1.05 & 0.29 & 0.08 & 0.39 & 0.29 & 0.08 & 0.39 & 0.29 & 0.08 \\ 
   &  & NAIVE & -43.59 & 0.20 & 0.23 & -43.59 & 0.20 & 0.23 & 21.73 & 0.27 & 0.12 & 21.73 & 0.27 & 0.12 \\ 
   &  & 25 & 2.54 & 1.81 & 3.28 & 5.93 & 1.56 & 2.43 & -1.93 & 0.88 & 0.78 & -3.35 & 0.79 & 0.63 \\ 
   &  & 50 & 5.36 & 0.63 & 0.40 & 5.02 & 0.64 & 0.41 & -2.82 & 0.41 & 0.17 & -2.64 & 0.41 & 0.17 \\ 
   &  & 100 & 2.27 & 0.52 & 0.27 & 2.33 & 0.53 & 0.28 & -1.33 & 0.38 & 0.14 & -1.43 & 0.38 & 0.15 \\ 
   &  & 200 & 0.31 & 0.44 & 0.19 & 0.47 & 0.45 & 0.20 & -0.04 & 0.35 & 0.12 & -0.21 & 0.35 & 0.12 \\ 
   &  & 400 & -0.33 & 0.38 & 0.15 & -0.22 & 0.38 & 0.15 & 0.08 & 0.32 & 0.10 & -0.08 & 0.32 & 0.10 \\ 
   & 0 & TRUE & -1.05 & 0.29 & 0.08 & -1.05 & 0.29 & 0.08 & 0.39 & 0.29 & 0.08 & 0.39 & 0.29 & 0.08 \\ 
   &  & NAIVE & -57.78 & 0.20 & 0.37 & -57.78 & 0.2 & 0.37 & 28.85 & 0.27 & 0.16 & 28.85 & 0.27 & 0.16 \\ 
   &  & 25 & 25.03 & 3.66 & 13.48 & 25.41 & 3.61 & 13.07 & -12.84 & 1.94 & 3.76 & -13.09 & 1.92 & 3.70 \\ 
   &  & 50 & 8.07 & 0.69 & 0.48 & 7.48 & 0.68 & 0.47 & -4.18 & 0.43 & 0.19 & -3.87 & 0.43 & 0.19 \\ 
   &  & 100 & 3.51 & 0.54 & 0.29 & 3.47 & 0.55 & 0.3 & -1.94 & 0.39 & 0.15 & -1.99 & 0.39 & 0.15 \\ 
   &  & 200 & 0.81 & 0.45 & 0.20 & 0.93 & 0.46 & 0.21 & -0.29 & 0.35 & 0.13 & -0.42 & 0.36 & 0.13 \\ 
   &  & 400 & -0.16 & 0.39 & 0.15 & -0.04 & 0.39 & 0.15 & -0.01 & 0.32 & 0.10 & -0.17 & 0.32 & 0.10 \\ 
  0 & 0.5 & TRUE & -0.91 & 0.24 & 0.06 & -0.91 & 0.24 & 0.06 & 0.10 & 0.25 & 0.06 & 0.10 & 0.25 & 0.06 \\ 
   &  & NAIVE & -25.76 & 0.18 & 0.10 & -25.76 & 0.18 & 0.10 & 0.14 & 0.25 & 0.06 & 0.14 & 0.25 & 0.06 \\ 
   &  & 25 & 3.91 & 0.62 & 0.39 & 3.35 & 0.59 & 0.35 & 0.17 & 0.26 & 0.07 & 0.18 & 0.26 & 0.07 \\ 
   &  & 50 & 3.15 & 0.83 & 0.70 & 3.00 & 0.83 & 0.69 & 0.00 & 0.27 & 0.07 & 0.01 & 0.26 & 0.07 \\ 
   &  & 100 & 0.25 & 0.38 & 0.14 & 0.34 & 0.39 & 0.15 & 0.03 & 0.27 & 0.07 & -0.08 & 0.27 & 0.07 \\ 
   &  & 200 & -0.24 & 0.35 & 0.12 & -0.11 & 0.36 & 0.13 & 0.24 & 0.27 & 0.07 & 0.17 & 0.27 & 0.07 \\ 
   &  & 400 & -0.5 & 0.31 & 0.09 & -0.47 & 0.31 & 0.09 & 0.15 & 0.25 & 0.06 & 0.03 & 0.25 & 0.06 \\ 
   & 0.25 & TRUE & -0.91 & 0.24 & 0.06 & -0.91 & 0.24 & 0.06 & 0.10 & 0.25 & 0.06 & 0.10 & 0.25 & 0.06 \\ 
   &  & NAIVE & -38.18 & 0.18 & 0.18 & -38.18 & 0.18 & 0.18 & 0.14 & 0.25 & 0.06 & 0.14 & 0.25 & 0.06 \\ 
   &  & 25 & 12.14 & 1.79 & 3.23 & 10.71 & 1.53 & 2.34 & 0.45 & 0.28 & 0.08 & 0.42 & 0.27 & 0.07 \\ 
   &  & 50 & -2.03 & 1.44 & 2.06 & -2.17 & 1.42 & 2.01 & 0.21 & 0.27 & 0.07 & 0.22 & 0.27 & 0.07 \\ 
   &  & 100 & 0.97 & 0.39 & 0.15 & 1.01 & 0.4 & 0.16 & 0.04 & 0.27 & 0.07 & -0.07 & 0.27 & 0.07 \\ 
   &  & 200 & 0.07 & 0.35 & 0.12 & 0.19 & 0.36 & 0.13 & 0.24 & 0.27 & 0.07 & 0.16 & 0.27 & 0.07 \\ 
   &  & 400 & -0.38 & 0.31 & 0.10 & -0.35 & 0.31 & 0.10 & 0.15 & 0.25 & 0.06 & 0.02 & 0.25 & 0.06 \\ 
   & 0 & TRUE & -0.91 & 0.24 & 0.06 & -0.91 & 0.24 & 0.06 & 0.10 & 0.25 & 0.06 & 0.10 & 0.25 & 0.06 \\ 
   &  & NAIVE & -50.59 & 0.18 & 0.29 & -50.59 & 0.18 & 0.29 & 0.14 & 0.26 & 0.07 & 0.14 & 0.26 & 0.07 \\ 
   &  & 25 & 3.76 & 3.24 & 10.5 & 4.43 & 2.82 & 7.93 & 0.20 & 0.30 & 0.09 & 0.24 & 0.29 & 0.09 \\ 
   &  & 50 & 1.04 & 0.98 & 0.97 & 0.78 & 0.97 & 0.95 & 0.17 & 0.27 & 0.07 & 0.19 & 0.27 & 0.07 \\ 
   &  & 100 & 1.67 & 0.40 & 0.16 & 1.66 & 0.41 & 0.17 & 0.04 & 0.27 & 0.07 & -0.06 & 0.27 & 0.07 \\ 
   &  & 200 & 0.37 & 0.36 & 0.13 & 0.48 & 0.37 & 0.13 & 0.23 & 0.27 & 0.07 & 0.16 & 0.27 & 0.07 \\ 
   &  & 400 & -0.28 & 0.31 & 0.10 & -0.24 & 0.31 & 0.10 & 0.14 & 0.25 & 0.06 & 0.01 & 0.25 & 0.06 \\     \hline \\
    \end{tabular}
  \end{center}

\end{sidewaystable}

\clearpage
\setcounter{table}{7}

\clearpage
\begin{table}
  \renewcommand{\thetable}{S\arabic{table}}

\caption{For 1000 simulated data sets, the mean percent (\%) bias,  empirical standard
error (SE), average estimated standard error (ASE), mean squared error
(MSE), and coverage probability for the 95\% confidence intervals (CP) are given for $(\beta_x,\beta_z)$, for linear regression of  energy on protein density (log-scale) and BMI using simulated true data (TRUE), the unadjusted self-reported data (Naive), and the proposed regression calibration method (Proposed). For the proposed method, the SE and coverage probability were calculated with using both the bootstrapped standard error (BSE) the sandwich estimator (SSE).
}
  \centering
\vspace{.1 in}

\begin{tabular}{llcccccccccccccc}
\hline
\hline
$ \beta_x$ & Method & \% Bias & SE & MSEx100 & ASE & CP \\ 
& TRUE & 0.0410 & 0.0062 & 0.0039 & 0.0064 & 0.957  \\ 
  & NAIVE & 16.238 & 0.0099 & 0.1069 & 0.0099 & 0.114  \\ 
  & Proposed BSE& -1.2704 & 0.0853 & 0.7278 & 0.0806 & 0.934  \\
    & Proposed SSE & -1.2704 & 0.0853 & 0.7278 & 0.0800 & 0.937 \\ \\

  $\beta_z$ & Method & \% Bias & SE & MSEx100 & ASE & CP \\ 
  & TRUE & -0.0106 & 0.0002 & 0.0000 & 0.0002 & 0.951 \\ 
 & NAIVE & -80.1324 & 0.0004 & 0.0110 & 0.0004 & 0.000  \\ 
  & Proposed BSE & 0.2963 & 0.0025 & 0.0006 & 0.0025 & 0.944  \\ 
    & Proposed SSE & 0.2963 & 0.0025 & 0.0006  & 0.0025 & 0.944 \\ 

  \hline
  \end{tabular}

\end{table}

\clearpage

\begin{figure}[hbt!]
  \renewcommand{\thefigure}{S\arabic{figure}}
\begin{center}
\caption{{Error distributions considered for the marginals for $T$ and $\tilde T$ for Table 3 (normal distribution) and Table S5 (mixture of two normal distributions and a log-normal distribution).}}
\includegraphics[width=6.5in]{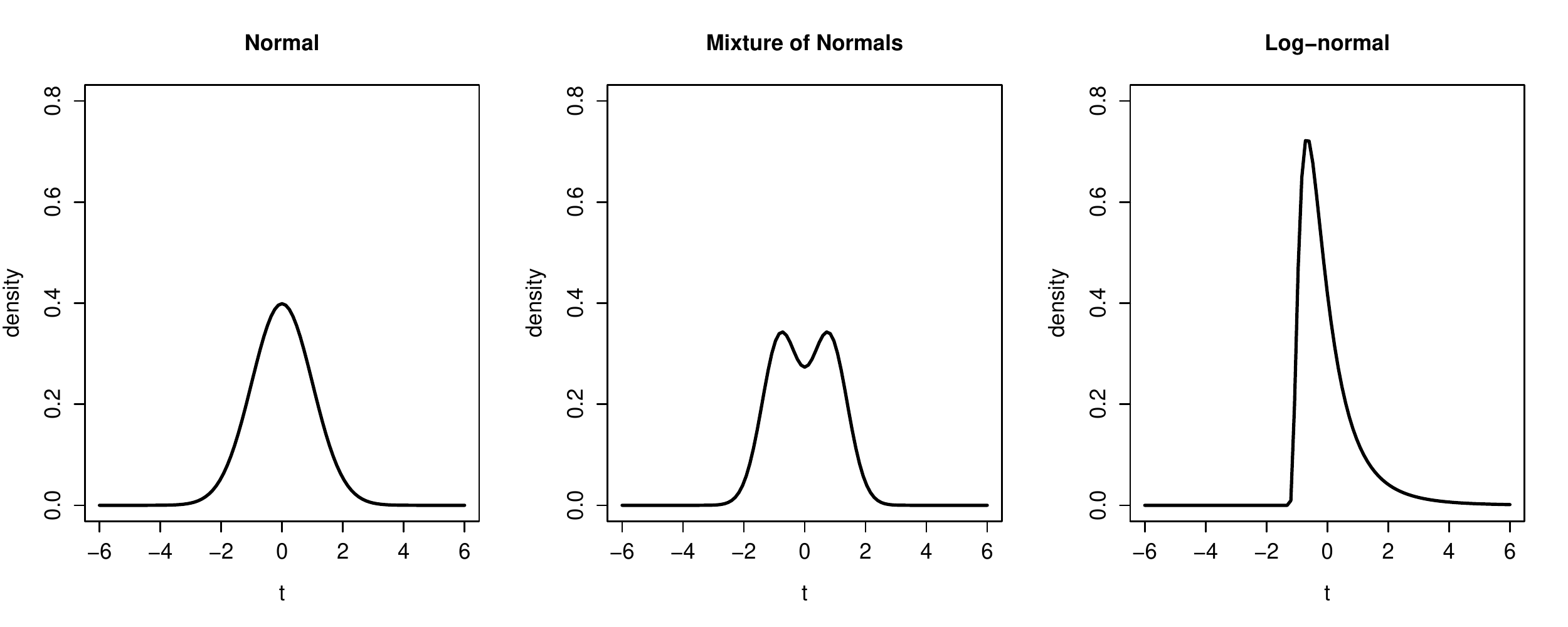}
\label{Figure1}
\end{center}
\end{figure}

\clearpage 

\subsection{Simulation Details for WHI Data Example}

In Section 5 of the manuscript, we performed a simulation study based on data from the Women's Health Initiative's Dietary Modification (DM) Trial and Nutritional Biomarker Substudy (NBS). The true model for the data was assumed be as in case 3 in Section 2.3.  Dietary nutrients were analyzed on the log-scale and assumed log-normal.  We generated the true log protein density intake $(X)$ and BMI $(Z)$ from a bivariate normal distribution with mean $(\mu_X,\mu_Z) = (2.647, 28.228)$ and covariance $(\sigma_X= 0.199, \sigma_Z = 5.547,\rho_{XZ} =  0.0043)$. The true log-energy intake (Y) was generated by the model $Y=\beta_0 + \beta_1 X + \beta_z Z+\epsilon$, where $(\beta_0,\beta_X,\beta_z)= (7.76, -0.192 , 0.013$) and $\epsilon$ a normally distributed random error with mean 0 and standard deviation $\sigma_{\epsilon} = 0.101$. $U$ and $\tilde U$, the random errors terms in self-reported $X^*$ and $Y^*$, were generated from a bivariate normal distribution with mean $(0, 0)$ and covariance $(\sigma_{U_s} = 0.112, \sigma_{\tilde U_s} = 0.3, \rho_{U_s, \tilde U_s} = -0.12)$.  $X^*$ was generated from $X + \alpha_0 + \alpha_1 Z + U_s$, where $\alpha_0 = 0.207, \alpha_1 = 0.0036$. $Y^*$ was generated from $Y + \gamma_0 + \gamma_1 Z + \tilde U_s$, where $\gamma_0 = 0.0054, \gamma_1 = -0.0113$. $U_b$ and $\tilde U_b$, the classical errors in biomarkers $X_b$ and $Y_b$, were generated from a bivariate normal distribution with mean (0,0) and covariance $(\sigma_{U_b} = 0.186, \sigma_{\tilde U_b} = 0.084, \rho_{U_b, \tilde U_b} = 0)$.  The biomarkers were then generated assuming $X_b=X + U_b$ and $Y_b=Y + \tilde U_b$. 

The parameter were estimated from the WHI data as follows. We observed self-reported samples from $N = 29,562$ subjects from the DM cohort. There were $n = 544$ subjects in the NBS sub-cohort with biomarkers, with $n_r=103$ in a reliability subset with repeated measures of biomarker and self-reported values. Let $(X_{bij},Y_{b{ij}})$ and $(X^\star_{ij},Y^\star_{ij})$ denote the $j$-th measure for subject $i$ for the biomarker and self-reported protein and energy consumption, respectively. The following estimates are from the DM cohort: 
\begin{align*}
\hat{\mu}_{X^\star} = \frac{1}{N}\sum_{i=1}^N X^\star_{i}\\
\hat{\mu}_Z = \frac{1}{N}\sum_{i=1}^N Z_{i}.
\end{align*}
The following estimates are from the NBS reliability subset:
\begin{align*}
\hat{\sigma}_{U_s} = \sqrt{\mbox{var}(X^\star_{1} - X^{\star}_{2})/2} \\
\hat{\sigma}_{\tilde U_s} = \sqrt{\mbox{var}(Y^*_{1} - Y^*_{2})/2} \\
\hat{\sigma}_{\tilde U_b} = \sqrt{\mbox{var}(Y_{b1} - Y_{b2})/2} \\
\hat{\sigma}_{U_b} = \sqrt{\mbox{var}(X_{b1} - X_{b2})/2}. 
\end{align*}
The following estimates are from NBS biomarker subset:
\begin{align*}
\hat{\sigma}_X = \sqrt{\mbox{var}(X^\star) - \hat{\sigma}_{U_b}^2}\\
\hat{\sigma}_{XZ} = \sqrt{\mbox{cov}(X_b,Z)}\\
\hat{\sigma}_{U_s \tilde U_s} = \widehat{\mbox{cov}}(U_s,\tilde U_s) = \mbox{cov}(Y^* - Y_b - \hat{\gamma}_0 - \hat{\gamma}_1 Z, X^* - X_b - \hat{\alpha}_0 - \hat{\alpha}_1 Z)\\
\hat{\sigma}_{\epsilon} = \sqrt{(1-R^2) \{\mbox{var}(Y_b) - \hat{\sigma}_{\tilde U_b}^2\}},
\end{align*}
where $R^2$ is the corrected R-squared coefficient for the regression of the unobserved $Y$ on $(X,Z)$. $R^2$ is calculated by $R^{2*}/\lambda_Y$, where $R^{2*}$ is the R-squared coefficient from the regression of $Y_b$ on $(X_b,Z)$ and dividing by the biomarker reliability coefficient  $\lambda_Y = (\mbox{var}({Y_b}) - \hat{\sigma}_{\tilde U_b}^2)/\mbox{var}({Y_b})$ (Yasmin \textit{et al.} 2013). 
 The estimates for $\gamma_0, \gamma_1$ are obtained from the coefficients of regressing $Y^* - Y_b$ on $Z$. The estimates for $\alpha_0, \alpha_1$ are obtained from  the coefficients of regressing $X^* - X_b$ on $Z$. The estimates for $\beta_0, \beta_x, \beta_z$ are obtained from the coefficients of regression calibration with repeated measures of biomarker values. \\

\noindent {\bf References}
\begin{enumerate}
\item Mossavar-Rahmani, Y., Tinker, L. F., Huang, Y., Neuhouser, M. L., McCann, S. E., Seguin, R. A., Vitolins, M. Z., Curb, J.D., and Prentice, R. L. (2013). {\it Nutritional Journal} \textbf{12}:63, 1-14.
\end{enumerate}

\label{lastpage}
\end{appendices}

\end{document}